\documentclass[12pt]{article}
\usepackage{graphicx, eurosym, indentfirst,float,latexsym}
\usepackage{amsmath,amsthm,amsopn,amstext,amscd,amsfonts,amssymb}
\usepackage{pdflscape,color}
\usepackage[T1]{fontenc}
\usepackage[utf8]{inputenc}
\usepackage[spanish, english]{babel}
\def\build#1#2#3{\mathrel{\mathop{#1}\limits^{#2}_{#3}}}
\begin{document}
%--------------------------------------------------------------------
\title{The Hubbert diffusion process: estimation via simulated annealing and variable neighborhood search procedures. Application to forecasting peak oil production}
\author{Istoni da Luz Sant'Ana\thanks{Department of Mathematics. Federal Institute of Rio de Janeiro, Brazil.}\, , Patricia Román Román\thanks{Department of Statistics and Operations Research, University of Granada, Spain}\, and Francisco Torres-Ruiz \thanks{(corresponding author). Department of Statistics and Operations Research, University of Granada, Spain. Address: Departamento de Estad{\'\i}stica e I.O., Facultad de Ciencias. Avda. Fuentenueva s/n. 18071, Granada, Spain. Email: fdeasis@ugr.es}}
\date{}
\maketitle
%--------------------------------------------------------------------

\begin{abstract}
Accurately charting the progress of oil production is a problem of great current interest. Oil production is widely known to be cyclical: in any given system, after it reaches its peak a decline will begin. With this in mind, M.K. Hubbert developed his peak theory in 1956, based on the bell-shaped curve that bears his name. In the present work we consider a stochastic model, based on the theory of diffusion processes and associated with the Hubbert curve. The problem of the maximum likelihood estimation of the parameters for this process is also considered. Since a complex system of equations appears, with a solution that can not be guaranteed by classical numerical procedures, we suggest the use of metaheuristic optimization algorithms such as simulated annealing and variable neighborhood search. Some strategies are suggested for bounding the space of solutions, and a description is provided for the application of the algorithms selected. In the case of the
variable neighborhood
search algorithm, a hybrid method is proposed in which it is combined with simulated annealing. In order to validate the theory developed here, we also carry out some studies based on simulated data, and consider two real crude oil production scenarios from Norway and Kazakhstan.
\end{abstract}

\noindent \textbf{Keywords:}
Hubbert curve, \, Oil production model,\, Diffusion processes,\, Simulated annealing,\, Variable neighborhood search.

\section{Introduction}
\label{intro}
For several decades now, the forecasting of oil production has been a problem of great interest given its fundamental role in the world's economy. In fact, the growth rate of the world economy is directly linked to oil consumption.

Given that petroleum is a non-renewable and finite resource, it is imperative that we are able to forecast future oil production, and to predict the precise time at which production will peak. According to the concept of peak oil, in any given country (and in the world as a whole) oil production rates will eventually reach, or may have already reached, a maximum. After that point, production rates will start to decline.

Given the undeniable fact that oil fuels the world's economy, reaching production peak has unavoidable implications for economic growth. Some possible consequences are a slowdown of economic growth, the need to resort to more efficient energy usage, and the development of alternative energy sources, among others. Many studies have dealt with the consequences of peak oil, and some examples of this are the archives of the Association for the Study of Peak Oil \cite{AsoPeak} or the so-called Hirsch Report, which provides a good overview on the issue of peak oil, its implications, and its possible mitigation \cite{Hir05}. \cite{Cur09} provides an analysis of policy responses to climate change and peak oil, while \cite{Lut12} analyzes the economic effects of peak oil.

The pioneer of historical evaluations on crude oil depletion was the geologist M. K. Hubbert, who in 1956 correctly estimated that oil production in the USA would peak around 1970. In recent years, many authors
have been stimulated by the economic and political implications of energy problems and have turned their attention to Hubbert’s theory, in an attempt to apply his analysis to other countries and thus forecast the evolution of
world oil production. Some experts conclude that production has already peaked \cite{bakht04,deffe05}, whereas others argue that the peak will occur soon; concretely, in \cite{Tow14} it is concluded that the peak is not likely to occur before 2018, and that this deadline will be further extended by rising oil prices and technology developments.

The debate concerning oil depletion has become broader, with an abundance of analyses and predictions about the date of peak oil. Many of these models refer to Hubbert’s approach, and try to extend and update it. We must note that, although the peak theory pioneered by Hubbert was implemented in the oil production context, the related depletion analysis can also be applied to other non-renewable resources such as natural gas \cite{soldo12}, phosphorus \cite{dery07}, and lithium \cite{vikstrom13}.

This paper considers a stochastic process, related to the Hubbert curve, to deal with oil production, and it is structured as follows: Section 2 is devoted to a brief review of some of the models proposed so far for modeling oil production and estimating peak and peak time. Section 3 describes how the Hubbert process is obtained from a general expression, and how the reparametrization of the logistic curve is carried out.
Section 4 deals with how the parameters of the model are estimated using maximum likelihood, and with the subsequent estimation of peak and peak time. The complexity of the likelihood system of equations leads to a direct estimation by maximizing the likelihood function. Section 5 deals with this problem through the application of simulated annealing (SA) and variable neighborhood search (VNS) algorithms. Firstly, a brief summary of the algorithms is provided, and then their adaptation to the problem at hand is presented. Some strategies are suggested for bounding the space of solutions, and a description is provided for the application of the algorithms selected. In the case of the VNS algorithm, a hybrid method is proposed in which it is combined with SA. In order to validate the methodology described, Section 6 describes a simulation study. Finally, in Section 7, we present two applications to real data by considering crude oil production data from Norway and Kazakhstan. These examples show the possibilities that
the process affords for the modelization of oil production, and help answer the question of when peak production can be expected.

\section{Brief summary of oil production models}

M. K. Hubbert was the first researcher who developed a theory for the study of oil production. In 1956 he applied his theory to crude oil production in the US Lower 48 states and correctly predicted that its peak would be reached around 1970 \cite{RefHubbert}. At first, he did not provide a functional form for his prediction, but instead fit past production to a bell-shaped curve in which the area under the curve was equal to his estimates of the amount of total oil available. Later, in 1959, he specified a functional form for the curve \cite{Hubbert59}. His starting point was the logistic curve, stating that cumulative production would follow a logistic curve, and thus that yearly production would follow the first derivative of the logistic curve (which was named, since that moment, the Hubbert's curve).

Probably, the leading proponent of Hubbert’s theories is K. Deffeyes, who has published several books on the subject, the latest in 2010, \cite{deffe10}. However, some aspects of these theories have also led other researchers to extend and/or modify his original model.

One such aspect is related to the fact that Hubbert provides a forecast with only one peak in oil production, which seems valid for countries with a large number of oil fields and basins. In \cite{laher00} examples are presented showing that oil production in several countries (France, UK, and others) cannot be represented by a single Hubbert cycle, and a model characterized by several cycles is introduced. This approach has been labelled \textit{multiple-Hubbert modeling}, and has been extended in \cite{maggio09}, and applied to world oil production \cite{Nas10} and to oil production in Brazil \cite{Sar14}.

In addition, some authors have focused on the symmetry of the curve and on modifying this aspect. For example, \cite{Hal04,Hal14} uses a modified version of the bell-shaped curve, which peaks at 60\% of ultimate production instead of the typical 50\%. This method implies an asymmetric shape of production and a steeper rate of decline than increase. Symmetric and asymmetric linear and exponential models are considered in \cite{Bra07}, where they are compared with the Hubbert curve.

Another widely discussed aspect is that Hubbert's method assumes that oil production is only time-dependent, and therefore does not take into account the effect of possible technological and/or economic factors. This has led to the modification of the model via the introduction of economic variables and econometric models. For example, \cite{Rey99,Rey02} included prices and costs in Hubbert’s model. In \cite{Kau91} the effect of geological, economic, and political factors on oil production in the US lower 48 states between 1947 and 1985 was analyzed with a method that combines curve fitting and econometric models. \cite{Bre12} provides a hybrid approach to the peak-oil question with two models in which the use of logistic curves for cumulative production is supplemented with data on projected extraction costs and historical rates of capacity increase. Also, an econometric model based on a system of simultaneous equations was
developed in \cite{
Kem03}. Other economic approaches to this subject have considered models in which production increases with demand, advancing technology, reserve additions, and site development \cite{holla08}. In \cite{guseo07}, a generalized Bass model was introduced to treat global oil growth as a natural diffusion process linked to exogenous variables like price, technology, and strategic interventions.

As regards probabilistic modeling of oil production, the literature contains few references to the subject. In \cite{bertr11} a probabilistic model to predict oil production was suggested based on field size (modeled by a Pareto distribution), which takes into account the process of launching production (modeled by a gamma distribution).

\section{The Hubbert diffusion process}
\subsection{The Hubbert curve}

It is well known that the Hubbert curve is obtained from the derivative of the logistic function, for which we consider the general expression:
\begin{equation}
\label{logistica}
l(t) = \frac{k}{\eta + \alpha^t}, \qquad  t \in\mathbb{R};\, \eta,k> 0,\, \, 0<\alpha<1.
\end{equation}

By deriving (\ref{logistica}) with respect to $t$, and imposing that $l'(t_0)=x_0$ (here $t_0$ represents the initial observation time), the following expression for the Hubbert curve is obtained:
\begin{equation}
\label{HubbertCurve}
x(t) = l'(t) = x_0\left(\frac{\eta+\alpha^{t_0}}{\eta+\alpha^{t}}\right)^2\alpha^{t-t_0}, \; t\in \mathbb{R}; \,\eta> 0,\, \, 0<\alpha<1.
\end{equation}

In the context of oil production, $x(t)$ usually represents the number of barrels produced per day. Parameter $\eta$ has no units, whereas for $\alpha$ the units are $\exp(\mbox{days}^{-1})$.

The maximum of (\ref{HubbertCurve}) is, probably, the main feature of the Hubbert curve when it comes to modeling oil production. In fact, the maximum is known as the peak of production, and is achieved at time instant
\begin{equation}
\label{TiempoPico}
t_{max} = \ln \eta/ \ln \alpha,
\end{equation}
usually known as peak time. Furthermore, $t_{max} > t_0$, i.e. the peak occurs after the initial observation time, if and only if $0 < \eta < \alpha^{t_0} < 1$. In addition, its value is
\begin{equation}
\label{peak}
x\left(t_{max}\right) = x_0\frac{\left(\eta + \alpha^{t_0}\right)^2}{4\eta\alpha^{t_0}}\cdot
\end{equation}

Another important feature of the curve is given by its inflection points. It can be seen that the curve exhibits two inflection points, symmetric around $t_{max}$. Concretely,
\begin{equation*}
\label{tI}
\begin{array}{l}
   t_{inf,1} = t_{max} + \displaystyle\frac{\ln (2 + \sqrt{3})}{\ln \alpha},\\ \\
   t_{inf,2} = t_{max} + \displaystyle\frac{\ln (2 - \sqrt{3})}{\ln \alpha}\cdot
\end{array}
\end{equation*}
These points verify $t_{inf,1}<t_{max}<t_{inf,2}$. Furthermore, $t_{inf,1}>t_0$ if and only if $\eta < \alpha^{t_0}\left(2-\sqrt{3}\right) < 2-\sqrt{3}$.

Finally, we consider the area under the curve, known as ultimate recoverable resources (URR) in oil production and often used as a tool for estimating the parameters of the curve. Its expression is given by
\begin{equation}
\label{URR}
URR =  - x_0\frac{\left(\eta + \alpha^{t_0}\right)^2}{\eta\alpha^{t_0}\ln\alpha}\cdot
\end{equation}

\subsection{Obtaining the diffusion process}

In order to model Hubbert-type behaviors from a stochastic point of view, our contribution is to consider a diffusion process whose mean function is (\ref{HubbertCurve}). This expression verifies the ordinary differential equation
\begin{equation}
\label{edoH}
x'(t)=r(t)x(t), \ \ x(t_0)=x_0,
\end{equation}
where
$$
r(t)=-\ln\alpha\frac{\alpha^{t}-\eta}{\eta+\alpha^{t}},
$$
which can be viewed as a generalization of the Malthusian growth model with a time-dependent fertility rate $r(t)$. A stochastic version of this model is given by the nonhomogeneous lognormal diffusion process (or lognormal diffusion process with exogenous factors). This is a diffusion process $\{X(t); t\geq t_0\}$, taking values on $\mathbb{R}^+$ and with infinitesimal moments
\begin{equation*}
\label{momentos}
\begin{array}{l}
A_1(x,t)=h(t)x\\
A_2(x)={\sigma}^{2}x^{2},
\end{array}
\end{equation*}
being the solution of the stochastic differential equation
\begin{equation*}
\label{ede}
\begin{array}{l}
dX(t) =h(t)X(t) dt+ \sigma X(t) dW(t) \\
X(t_0)=X_0,
\end{array}
\end{equation*}
where $W(t)$ is a standard Wiener process independent on $X_0, \ t \geq t_0$ and $h(t)$ is a continuous and bounded function.

An explanation of the main features of the process can be found in \cite{Gut06}, where the authors made a detailed theoretical analysis of the process. Concerning other potential fields of application, in \cite{Gut99} an inferential analysis is performed to assess the usefulness of the process in Economics.

Among the characteristics of the process we will focus on the mean function and its conditioned version on a value $y$ in a previous time instant $s$, whose expressions are

\begin{equation}
\label{medias}
\begin{array}{l}
m_X(t) = E[X(t)]= E[X_0]\exp \left(\displaystyle\int_{t_0}^t h(u) d u \right) \\
m_X(t|y,s)= E[X(t)|X(s)=y]=y \, \exp \left( \displaystyle\int_{s}^t h(u) d u \right).
\end{array}
\end{equation}

These functions verify $m_X'(t)=m_X(t)h(t)$ and $m_X'(t|y,s)= m_X(t|y,s) h(t)$, that is, the same ordinary differential equation (\ref{edoH}) verified by the Hubbert curve (\ref{HubbertCurve}). This question leads us to define the Hubbert diffusion process as the particular case of the nonhomogeneous lognormal diffusion process considering $h(t)=r(t)$\footnote{Note that $r(t)$ is a decreasing continuous function verifying $-\ln\alpha<r(t)<\ln\alpha$, $\forall t \in \mathbb{R}$.}.

In this way, all the features of the Hubbert process can be obtained from those established in \cite{Gut06} for the lognormal diffusion process with exogenous factors. In particular, the transition probability density function (pdf) is, for $s<t$,

\begin{equation}
\label{pdf}
f(x,t|y,s) = \frac{1}{\sigma x \sqrt{2\pi(t-s)}} exp\left\lbrace\frac{-\left[\ln\displaystyle\frac{x}{y} - 2\ln\frac{\eta+\alpha^s}{\eta+\alpha^t}- \left(\ln\alpha - \frac{\sigma^2}{2}\right)(t-s)\right]^2}{2\sigma^2(t-s)}\right\rbrace,
\end{equation}
which corresponds to that of a lognormal variable, i.e.

\begin{align*}
%\label{fdt}
& X(t)|X(s) = y \sim \Lambda_1\left[\ln y + 2\ln \frac{\eta+\alpha^s}{\eta+\alpha^t} + \left(\ln\alpha - \frac{\sigma^2}{2}\right)(t-s), \sigma^2(t-s)\right].
\end{align*}

In order to calculate the finite-dimensional distributions of this process, the distribution of $X_0$ must be fixed. By considering a degenerate distribution at $t_0$, i.e. $P[X_0=x_0]=1$, or a lognormal distribution $\Lambda_1(\mu_0, \sigma_0^2)$, all finite-dimensional distributions are lognormal\footnote{Note that the former case is a particular case of the second, with $\mu_0= \log x_0$ and $\sigma_0^2=0$.}. Concretely, $\forall n \in \mathbb{N}$ and $t_1<t_2< \ldots < t_n$,
and denoting by $\Lambda_n$ the $n$-dimensional lognormal distribution, we have

\[
(X(t_1), \ldots, X(t_n))^T \ \sim \ \Lambda_n(\mu, \Sigma),
\]
where the components of vector $\mu=(\mu_1, \ldots, \mu_n)^T$ and matrix $\Sigma = (\sigma_{ij}), \ i, j=1, \ldots, n$, are

\[
\mu_i = \mu_0+ 2\ln\frac{\eta+\alpha^{t_0}}{\eta+\alpha^{t_i}}+\left(\ln\alpha - \frac{\sigma^2}{2}\right) (t_i-t_0), \ \ i= 1, \ldots, n
\]
and
\[
\sigma_{ij}= \sigma_0^2+ \sigma^2 (Min(t_i,t_j)-t_0), \ \ i, j=1, \ldots, n,
\]
respectively. Finally, from (\ref{medias}), the mean functions of the Hubbert process result in the forms
\begin{equation}
\label{media}
m_X(t)= E[X_0]\left(\frac{\eta+\alpha^{t_0}}{\eta+\alpha^t}\right)^2 \alpha^{(t-t_0)}, \,\, t\geq t_0
\end{equation}
and
\begin{equation}
\label{mediacondicionada}
m_X(t|y,s) = y\left(\frac{\eta+\alpha^{s}}{\eta+\alpha^t}\right)^2\alpha^{(t-s)}, \,\, t\geq s
\end{equation}
respectively, which are Hubbert curves of the type here introduced and can be used for predictions purposes within the context of this model.

\section{Inference on the process}
\label{inference}
In the context of oil production the prediction of peak and peak time is a problem of great interest. Furthermore, obtaining future production values can be very useful in real situations. This last question can be addressed using the mean functions (\ref{media}) and (\ref{mediacondicionada}). Note that these functions, as well as the peak time and peak given by (\ref{TiempoPico}) and (\ref{peak}) respectively, are functions expressed in terms of the parameters of the process. The parameters of the process must therefore be previously estimated if we intend to make estimations in real-life situations.

Let us then examine in this section the ML estimation of the parameters of the model, from which we can obtain, by virtue of Zehna's theorem\footnote{Zehna's theorem states that if $\widehat{\theta}$ is a ML estimator for $\theta$, then $g(\widehat{\theta})$ is a ML estimator for $g(\theta)$ (see \cite{Roa01}).}, the corresponding for the above-mentioned parametric functions.

\subsection{Likelihood function}

Let us consider a discrete sampling of the process, based on $d$ sample paths, at times $t_{ij}$, $(i = 1,\cdots, d, \ j=1, \cdots, n_i)$ with $t_{i1}$ = $t_1$, $i = 1$, $\cdots$, $d$. Denote by $\mathbf{x}=\left\lbrace x_{ij}\right\rbrace_{i = 1, \cdots, d; j = 1, \cdots, n_i}$ the observed values of the $X\left(t_{ij}\right)$ variables of the process.

The likelihood function depends on the choice of the initial distribution. In the following we will consider the general case when the initial distribution is lognormal, i.e. $X(t_1)\sim\Lambda_1\left(\mu_1,\sigma^2_1\right)$. The transition pdf \eqref{pdf} can be rewritten as

\begin{align*}
 f(x_{ij},t_{ij}|x_{ij-1},t_{i,j-1}) &= \frac{1}{\sigma x_{ij} \sqrt{2\pi(t_{ij}-t_{i,j-1})}}  \nonumber \\
 &\times\exp\left(-\displaystyle\frac{\left[\ln\left(\frac{x_{ij}}{x_{i,j-1}}\right) - 2T_{ij}^{\eta,\alpha}-\left(\ln\alpha -\frac{\sigma^2}{2}\right)(t_{ij}-t_{i,j-1})\right]^2}{2\sigma^2(t_{ij}-t_{i,j-1})}\right)
\end{align*}
where
$$
T_{ij}^{\eta,\alpha} = \ln\frac{\eta + \alpha^{t_{i,j-1}}}{\eta + \alpha^{t_{ij}}},
$$
from which, and denoting $N = \sum_{i=1}^d n_i$, the log-likelihood function of the sample is

\begin{align}
\label{vero}
L_{\mathbf{x}}\left(\mu_1,\sigma^2_1,\eta,\alpha,\sigma^2\right)&=-\frac{N}{2}\ln(2\pi) - \frac{d}{2}\ln\sigma^2_1-\frac{N-d}{2}\ln\sigma^2-\sum_{i=1}^{d}\ln x_{i1}\nonumber \\ &-\frac{1}{2\sigma^2_1}\sum^d_{i=1}\left[\ln x_{i1}-\mu_1\right]^2-\sum_{i=1}^{d}\sum_{j=2}^{n_i}\ln x_{ij} - \frac{1}{2}\sum_{i=1}^{d}\sum_{j=2}^{n_i}\ln\left(t_{ij}-t_{i,j-1}\right)\nonumber\\
&- \frac{1}{2\sigma^2}\left[Z_1+4\left(Y_1^{\eta,\alpha}-Y_2^{\eta,\alpha}\right)+\left(\ln\alpha-
\frac{\sigma^2}{2}\right)\left[\left(\ln\alpha-\frac{\sigma^2}{2}\right)Z_2\right.\right.\nonumber\\
&\left.\left.-2\left[Z_3 - 2R^{\eta,\alpha}\right]\right]\right]
\end{align}
where

\begin{eqnarray*}
&Y_1^{\eta,\alpha} =\displaystyle \sum_{i = 1}^d \sum^{n_i}_{j=2} \frac{\left(T_{ij}^{\eta,\alpha}\right)^2}{t_{ij} - t_{i,j-1}},\  Y_2^{\eta,\alpha} = \displaystyle\sum_{i = 1}^d \sum^{n_i}_{j=2} \frac{\ln\left(\frac{x_{ij}}{x_{i,j-1}}\right)T_{ij}^{\eta,\alpha}}{t_{ij} - t_{i,j-1}},\ R^{\eta, \alpha}= \sum_{i=1}^d \ln \frac{\eta + \alpha^{t_{i1}}}{\eta + \alpha^{t_{in_{i}}}} \\
&Z_1 =\displaystyle\sum_{i = 1}^d \sum^{n_i}_{j=2} \frac{\ln^2\left(\frac{x_{ij}}{x_{i,j-1}}\right)}{t_{ij} - t_{i,j-1}}, \ Z_2 = \displaystyle\sum_{i = 1}^d\left(t_{in_i} - t_{i1}\right),\ Z_3 = \displaystyle\sum_{i = 1}^d \ln\left(\frac{x_{in_i}}{x_{i1}}\right).
\end{eqnarray*}

\subsection{Obtaining the ML estimates}

From (\ref{vero}) the ML estimates of $\mu_1$ and $\sigma_1^2$ are
\[
  \widehat{\mu}_1=\displaystyle\frac{1}{d}\displaystyle\sum_{i=1}^{d}\ln x_{i1} \ \ \mbox{and} \ \
  \widehat{\sigma}_1^2=\displaystyle\frac{1}{d}\displaystyle\sum_{i=1}^{d}(\ln x_{i1}-\widehat{\mu}_1)^2.
\]

However, estimating $\eta$, $\alpha$, and $\sigma^2$ poses some difficulties. In fact, the resulting system of equations is exceedingly complex and does not have an explicit solution, and numerical procedures must be employed. Nevertheless, it is impossible to carry out a general study of the system of equations in order to check the conditions of convergence of the chosen numerical method, since it is dependent on sample data.

One alternative would be using stochastic optimization procedures like SA and VNS. These algorithms are designed to solve problems of the type $\min_{\theta\in\Theta}g(\theta)$, being $g$ the target function to be optimized, and are often more appropriate than classical numerical methods since they impose fewer restrictions on the space of solutions $\Theta$ and on the analytical properties of $g$.

In our case, once $\mu_1$ and $\sigma_1^2$ have been estimated, the problem becomes maximizing function
$L_{\mathbf{x}}(\widehat{\mu}_1,\widehat{\sigma}_1^2,\eta,\alpha,\sigma^2)$. Since the algorithms mentioned above are usually formulated for minimization problems, from (\ref{vero}) the target function we will consider is
\begin{align}
\label{fobjetivo}
g_{\mathbf{x}}(\eta,\alpha,\sigma^2)&=\frac{N-d}{2}\ln\sigma^2
+\frac{1}{2\sigma^2}\left[Z_1+4\left(Y_1^{\eta,\alpha}-Y_2^{\eta,\alpha}\right)+
\left(\ln\alpha-\frac{\sigma^2}{2}\right)
\left[\left(\ln\alpha-\frac{\sigma^2}{2}\right)Z_2\right.\right.\nonumber\\
&\left.\left.-2\left[Z_3 - 2R^{\eta,\alpha}\right]\right]\right]
\end{align}

\subsection{Fitting and forecasting. Peak time and peak.}

Fitting the observed data, as well as predicting the future behavior of the process, can be done from the ML estimates of mean functions (\ref{media}) and (\ref{mediacondicionada}). Note that in the case of calculating predictions at future time instants, it is more reasonable to use the conditioned version of the mean function since it employs more updated information than that provided by the initial distribution of the process.

In addition, and given that the mean functions of the process are Hubbert functions,
forecasts for peak time can be performed by means of point estimation, substituting the ML estimates of the parameters in the time instant at which their maxima are achieved. We note that this time instant is the same in both cases and coincides with (\ref{TiempoPico}). Therefore, the ML estimation of peak time is
$$
\widehat{t}_{max}=\frac{\ln\widehat{\eta}}{\ln\widehat{\alpha}}\cdot
$$

The estimation of the peak is obtained by substituting that of peak time in (\ref{media}) or (\ref{mediacondicionada}). Note that we have two possibilities depending on the mean function chosen, resulting in
\begin{equation}
\label{pico1}
\widehat{E[X(t_1)]}\frac{\left(\widehat{\eta}+\widehat{\alpha}^{t_1}\right)^2}{4\widehat{\eta}\widehat{\alpha}^{t_1}},
\end{equation}
which coincides with (\ref{peak}) for a degenerate initial distribution, and
\begin{equation}
\label{pico2}
y\frac{\left(\widehat{\eta}+\widehat{\alpha}^{s}\right)^2}{4\widehat{\eta}\widehat{\alpha}^{s}},
\end{equation}
when the conditioned version is used.

At this point an important remark must be made about the values of time instants. In fact, when dealing with high values for the time instants of the process, it is possible that the estimated value of $\eta$ is close to zero (note that $\eta=\alpha^{t_{max}}$). In these situations, common in real cases, the accuracy in the estimates can cause detrimental effects on the final results. One way to tackle this problem is to consider a new diffusion process $\{Y(t); t\geq t_0-k\}$ obtained from $\{X(t); t\geq t_0\}$ by considering a shift of length $k$ in time, that is $Y(t)=X(t+k)$. It is not difficult to see that the new process is also a Hubbert diffusion process whose parameters $\alpha$ and $\sigma$ remain invariant while $\eta$ becomes $\eta'=\alpha^{-k}\eta$. Furthermore, the expressions of the mean functions, as well as the peak, are the same whether we use $\eta$ or $\eta'$, while for peak time it is enough to undo the changes made in time. In practice we recommend considering $k=t_0$ so that
the original data can be interpreted as observations of the new process with initial time instant $t_0=0$.

It is important to note that the value of the peak is independent of the use of $\eta$ or $\eta'$. In addition, if we note $m_Y(t|y,s)$ the conditioned mean function of process $Y(t)$, then
$m_Y(t|y,s)=m_X(t+x|y,s+k)$.

\section{Application of SA and VNS for estimating the parameters of the Hubbert diffusion process}

\subsection{A brief summary of SA and VNS}

SA is a local metaheuristic algorithm introduced in \cite{Kir83}, and inspired by the metallurgical process of annealing as studied in statistical mechanics. The algorithm performs an iterative exploration of solution space $\Theta$ searching for improvements on the value of the target function, say $g$, and intends to avoid an attraction towards local minima. Concretely, let $\theta$ be the solution for a given iteration, $\theta'$ a new value selected in a neighborhood of $\theta$ in the next iteration and
$\Delta=g(\theta')-g(\theta)$. If $\Delta\leq 0$ then $\theta'$ is selected as the new solution. Otherwise, it could be accepted with probability $p=\exp(-\Delta/T)$, where $T$ is a scale factor called \textit{temperature}. This selection procedure, usually referred as Metropolis algorithm, was described in \cite{Met53}.

When applying SA, temperature $T$ is gradually decreased in such a way that, at the beginning, the cooling process allows to select solutions which worsen the target function. As $T$ decreases such solutions are not longer accepted. Theoretical studies have shown that, during an infinitely slow cooling, the process converges (as $T$ approaches zero) to a global minimum with probability one.

Therefore, the application of SA requires:

\begin{enumerate}
\item Initializing the parameters of the algorithm: initial solution ($\theta_0$), initial and final temperature ($T_0$, $T_F$), chain length for each application of the Metropolis algorithm ($L$), cooling procedure, and a stopping condition.
\item Applying the Metropolis procedure $L$ times.
\item Checking the stopping rule and checking if the final temperature has been achieved. If both conditions are not met, decreasing $T$ and returning to the previous step.
\end{enumerate}

As regards the VNS algorithm introduced in \cite{Mla97}, its goal is to explore several neighborhoods in $\Theta$ when a local optimum is found for $g$ through a local search method. The algorithm is applied in two phases: in the first a structure of neighborhoods, say $N_k, k=1,\ldots,k_{max}$, is determined in the solution space and an initial solution $\theta_0$ is chosen. For $k=1,\ldots,k_{max}$, the second phase uses a local search method to determine a new $\theta^*$ solution in $N_k(\theta_0)$. If $\theta^*$ improves the target function, then $\theta_0=\theta^*$ and the search is recommenced from $N_1(\theta_0)$. If not, the search continues with $N_{k+1}(\theta_0)$. Note that the final solution is a local minimum with respect to the $k_{max}$ neighborhood structures, and therefore finding a global minimum is much more probable than when using a single structure. Also note that the procedure here described changes neighborhood every time there is an improvement in $g$. Another variations
exist depending on the different ways that the neighborhood structure may change when a local optimum has been reached, as well as on several local search methods (for other possibilities, see \cite{Bou13}).

Both algorithms have experienced remarkable popularity in the last years, having been applied to several fields of research. \cite{Ver08}, \cite{Abb11}, and \cite{Zor12} used them in the context of ML estimation for distributions, whereas in \cite{Rom12b} and \cite{Rom15} they helped estimate the parameters of Gompertz-type and Richards diffusion processes, respectively.

\subsection{Bounding the solution space}

The parametric space $\Theta$ linked to the target function (\ref{fobjetivo}), on which the selected algorithms must operate, is continuous and unbounded. Concretely
$$
\Theta=\{(\eta,\alpha,\sigma):\,\eta>0,\, 0<\alpha<1,\, \sigma>0\}.
$$
The drawback is that the solution space might not be explored with enough depth. This requires us to find arguments for bounding said space.

Regarding parameter $\sigma$, it is known that when it has high values it leads to sample paths with great variability around the mean of the process. Thus, excessive variability in available paths would make a Hubbert-type modeling inadvisable. Some simulations performed for several values of $\sigma$ have led us to consider that $0<\sigma<0\mbox{.}1$, so that we may have paths compatible with a Hubbert-type growth. On the other hand, whereas $\alpha$ is bounded, there does not seem to be an upper bound for $\eta$. However, as noted in Section 3, the Hubbert curve has an inflection point before $t_{max}$ which is visualized ($t_{inf}>t_0$) if and only if $\eta<\alpha^{t_0}(2-\sqrt{3})<2-\sqrt{3}$, so an upper bound for $\eta$ is found.

Additionally, when an estimation of the URR is available, some refinements can be made for bounding $\alpha$:
\begin{itemize}
  \item From (\ref{URR}) the following second degree equation is obtained
    \begin{equation*}
    \label{URR3}
    x_0\eta^2 + \alpha^{t_0}\left(2x_0 + URR\ln\alpha\right)\eta + x_0\alpha^{2t_0}=0.
    \end{equation*}
    In order to have a solution, the discriminant of the equation must verify
    $$
    \alpha^{2t_0}URR\ln\alpha\left(URR\ln\alpha+ 4x_0\right)\geq 0,
    $$
    from which we deduce $\alpha \leq \alpha_1$ where $\alpha_1=exp\left(-4x_0/URR\right)$.
  \item Now, let $T_{F}$ be the final observation time (it can be before or after $t_{inf}$ and/or $t_{max}$) and consider $c$ defined by
    \begin{equation*}
    \label{c/URR}
    c = \int_{t_0}^{T_F}x(t)dt = \eta URR\frac{\alpha^{t_0} - \alpha^{T_F}}{\left(\eta+\alpha^{t_0}\right)\left(\eta+\alpha^{T_F}\right)},
    \end{equation*}
    from which we obtain a new second degree equation, concretely
    \begin{equation*}
    \label{cURR2}
    M\eta^2 + \alpha^t_{0}\left[\left(M-1\right) + \left(M+1\right)\alpha^h\right]\eta + M\alpha^{2t_0 + h} = 0
    \end{equation*}
    where $h=T_F-t_0$ and $M=c/URR$.

    Following a similar argument to the one established for the previous bound, the following expression is now verified,
       $\left(M-1\right)^2 - 2\left(M^2+1\right)\alpha^{h}+ \left(M+1\right)^2\alpha^{2h} \geq 0$, resulting $\alpha \leq \alpha_2$ being
        $$
         \alpha_2 = \left(\frac{M-1}{M+1}\right)^{2/h}.
        $$
  \item From the previous bounds we consider $0<\alpha<\alpha^*=Min(\alpha_1,\alpha_2)$.
\end{itemize}

Table \ref{Bounds} contains, for several values of $\alpha$ and $\eta$, the bounds provided by the proposed method. We can see how for $\eta=0.01$ the bound provided by $\alpha_2$ is preferable, regardless of the value of $\alpha$. This situation changes as $\eta$ grows, since the range of values of $\alpha$ for which the bound provided by $\alpha_1$ is preferable increases. In such a case, as $\eta$ grows, the amplitude of the intervals provided by $\alpha_1$ decreases.

\subsection{Choosing the main options for applying SA and VNS}

Once the solution space has been bounded, we specify the choice of the initial parameters of the algorithms and the stopping conditions in order to apply them to the estimation of parameters in the Hubbert diffusion process.

For SA we consider:
\begin{enumerate}
   \item The initial solution is chosen randomly in the bounded subspace $\Theta^{\prime} = (0,2-\sqrt3)\times (0,\alpha^*)\times(0,0.1)$. Note that we have considered the shorter space, based on the previous bounding. This depends on the availability of data about URR. If these data are not available, then we must replace $\Theta'$ with $\Theta^{*} = (0,2-\sqrt3)\times (0,1)\times(0,0.1)$.
   \item For the application of SA the initial temperature must be high, so that at the beginning there is a high probability, $p_0$, of accepting values that increase the value of the target function. In our case, and following Busetti \cite{Bus03}, we have considered $p_0=0.9$, so that $T_0=-\Delta\,g^+/\log(p_0)$. $\Delta\,g^+$ denotes the average increase in the target function when values that produce an increase are accepted after considering $N$ values in the solution space. In our case, we have considered $N=100$.
   \item For the cooling process we have considered a geometric scheme in which the current temperature is multiplied by a constant $\gamma$ ($0<\gamma<1$), i.e. $T_i=\gamma\,T_{i-1}$, $i\geq 1$. The usual values for $\gamma$ are between 0.8 and 0.99. For this case we have set the constant at $\gamma=0\mbox{.}95$.
   \item The selected length of the chain for the application of the Metropolis algorithm is $L=50$. Therefore, in each step a chain of 50 solutions will be generated before checking the stopping rule and modifying the temperature if necessary.
   \item The selected stopping rule is twofold. Firstly, it checks whether the latest 50 values generated are equal, in which case the algorithm is stopped. Otherwise, it continues until the temperature reaches a value close to zero (0.1 in this case).
\end{enumerate}

As previously stated, for the application of the VNS algorithm we must take into account the neighborhood structure and the local search method. We have made the following choices:
\begin{itemize}
\item\textbf{Neighborhood structure.}
After selecting a value for $k_{max}$ (in the applications developed in this paper we have selected $k_{max}=5$) we proceed as follows.

Let $\Theta^{\prime}$ be the solution subspace mentioned above. Given an initial solution $\theta_0 = \left(\alpha_0,\eta_0,\sigma_0\right)$ we consider quantities
$$
\begin{array}{cccccc}
  h_{11} = \frac{\eta_0}{k_{max}},   & h_{12}=\frac{2-\sqrt3-\eta_0}{k_{max}}, &  h_{21} = \frac{\alpha_0}{k_{max}} & h_{22}=\frac{\alpha^*-\alpha_0}{k_{max}},
  h_{31} = \frac{\sigma_0}{k_{max}}, & h_{32} = \frac{0.1-\sigma_0}{k_{max}}
\end{array}
$$
from which the neighborhood structure is given by
\begin{align*}
N_{k}(\theta)&=[\eta_0-k\,h_{11},\eta_0+k\,h_{12}]\times[\alpha_0-k\,h_{21},\alpha_0+k\,h_{22}]
\times[\sigma_0-k\,h_{31},\sigma_0+k\,h_{32}]
\end{align*}
for $k=1,\ldots,k_{max}$.
\item \textbf{Local search.}
The local search method we have selected is SA. This choice allows us to perform a hybrid procedure which has proven to be useful in several applications (see, for instance \cite{Abb11} or \cite{Rom15}).
\item \textbf{Initial solution.} In order to apply the algorithm we have considered as an initial solution the one found by applying the SA algorithm.
\end{itemize}

\section{Simulation example}

In order to validate the procedures suggested in previous sections, we have performed a simulation study with the following pattern: 50 sample paths were simulated, each one obtained through the existing relation between the Hubbert process and the Wiener process. These sample paths are linked by the expression
$$
X(t)=X_{t_0}\left(\displaystyle\frac{\eta+\alpha^{t_0}}{\eta+\alpha^{t}}\right)^2\alpha^{t-t_0}\exp\left(\sigma W(t-t_0)-\displaystyle\frac{\sigma^2}{2}(t-t_0)\right).
$$

We have considered 501 data at time instants $t_i=(i-1)\cdot 0.1$, $i=1,\ldots,501$. We have proposed a set of values for parameters $\eta=(0.01, 0.025, 0.05, 0.1, 0.15, 0.2, 0.25)$ (note that $\eta<2-\sqrt{3}$), $\alpha=0.05+(i-1)\cdot 0.1$, $i=1,\ldots,10$, and $\sigma=(0.05,0.07)$, chosen arbitrarily. However, we wanted these values to be able to mirror real cases (as the ones described in the next section). The initial distribution is degenerate at value $x_0=100$. After the simulation, we chose a sample from the first value, using a step equal to 1. Hence, a sample of 51 data was obtained for each sample path. Furthermore, the whole procedure was replicated 50 times.

Table \ref{Errors} shows, for each combination of the parameters, the absolute relative errors ($\times 10^{-3}$) between the real and the estimated log-likelihood function after applying the SA and the hybrid VNS-SA algorithm. Note that both methods provide good estimates of the parameters in terms of the relative error in the likelihood function. Nevertheless, we must remark that using VNS with an initial solution given by SA improves the estimation of the parameters since a noticeable decrease in the relative error is observed. Finally, tables \ref{Estimates1} and \ref{Estimates2} show the estimates of the parameters provided by VNS.

\section{Real application}
\label{pratical}
In this section we consider real data, taken from the \textit{U.S. Energy Information Administration} \cite{iea}, concerning crude oil production (including lease condensate) from Norway and Kazakhstan. The aim is to fit a Hubbert diffusion process to both cases by applying the methodology developed in the previous sections, and then to obtain stochastic models showing the behavior of oil production in both countries. Thanks to these models we can:
\begin{itemize}
  \item Obtain the estimate mean function of the production.
  \item Make forecasts about oil production in the future.
  \item Estimate production peak time and peak.
\end{itemize}

\begin{figure}
\caption{Observed crude oil production data for Norway.}
\centering
\includegraphics[width=7cm, height = 6cm]{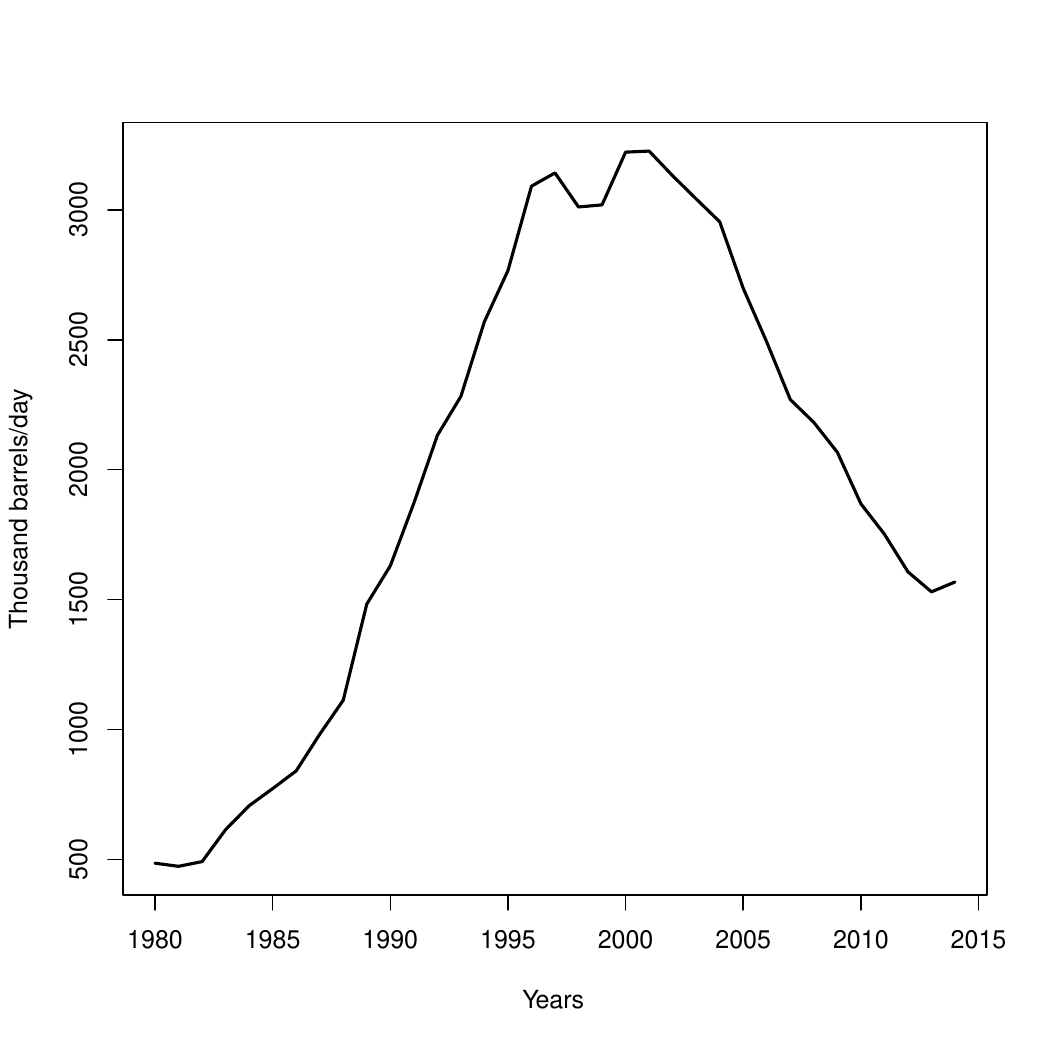}
\label{NorwayData}
\end{figure}

\begin{figure}
\caption{Observed crude oil production data for Kazakhstan.}
\centering
\includegraphics[width=7cm, height = 6cm]{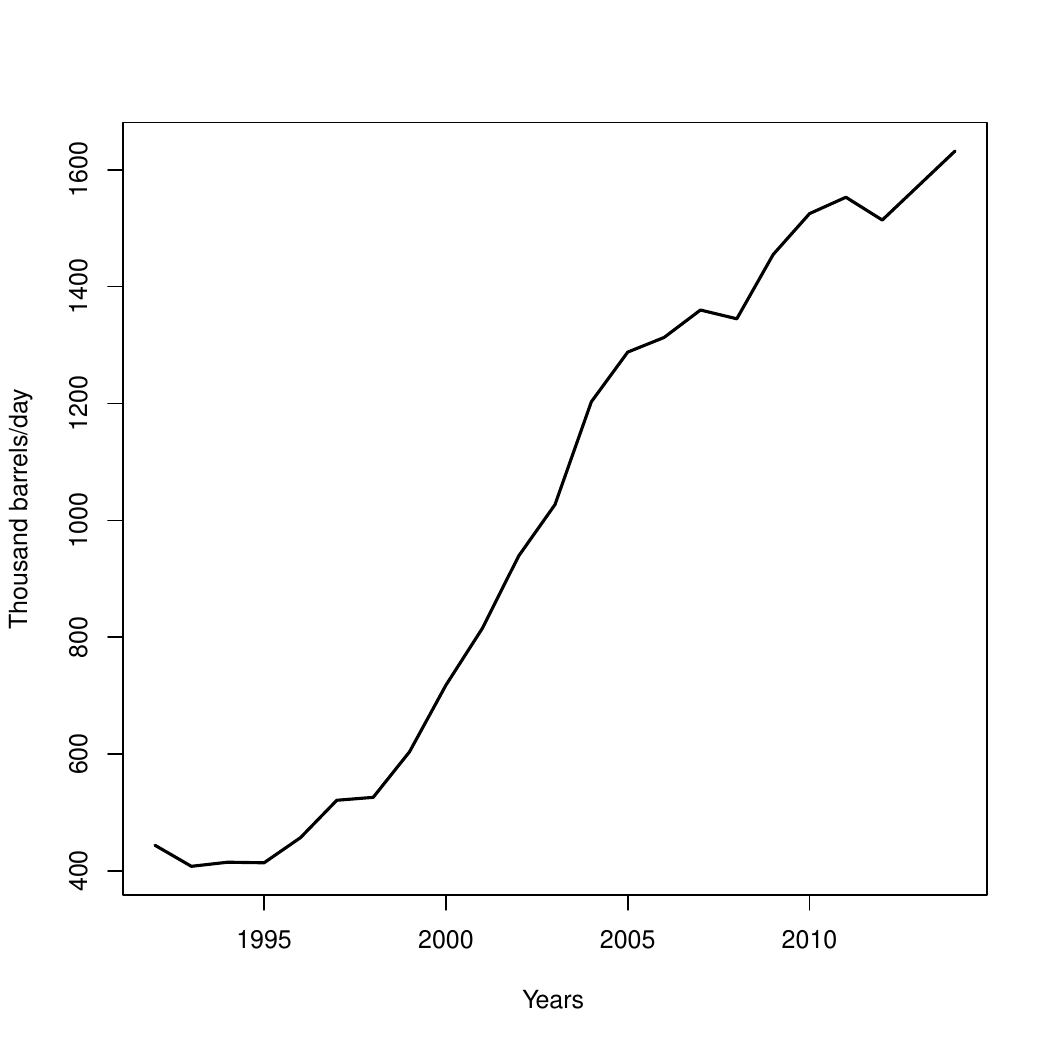}
\label{KazakhstanData}
\end{figure}

In the Norwegian case (Figure \ref{NorwayData}) the collected data range is 1980-2014, whereas for Kazakhstan (Figure \ref{KazakhstanData}) it is 1992-2014. In these graphs we observe that, in the case of Norway, production peak has already occurred, specifically in 2001 (with a production of 3226 thousand barrels per day (Mbbl/d)). After that year, its annual production started an uninterrupted decline with some exceptions such as 2014, when production increased by 2.5\% compared with 2013. This kind of behavior suggests that using the peak theory developed by Hubbert might be a good fit. A similar behavior may be expected for oil production in Kazakhstan if no exceptional factors influence its annual production, although its peak has not yet been reached. We will now analyze both cases, starting with Norway.

The data observed for Norway allows us to consider two different situations:

  \begin{itemize}
     \item \textbf{Scenario 1:} The peak is visualized. In this case, we have taken the whole observed sample path. This situation allows us to evaluate the capacity of the model for fitting the data.
     \item \textbf{Scenario 2:} The peak is not visualized. In this case we consider the truncated data of the previous one at a specific time $T_{F}$, before the visualization of the peak (in this case we have chosen $T_F=1999$). This scenario allows us to consider the predictive capabilities of the model to estimate peak time and to predict the future behavior of oil production.
  \end{itemize}

As we mentioned in Section 5, in order to find a bound for parameter $\alpha$ ( for all other parameters we have considered the fixed bounds mentioned in said section) we must have an estimation of the URR value. To this end, we have considered the sum of the estimated proved reserves, also collected by \cite{iea}, and the accumulated crude oil production until 2014. Table \ref{tab:norway} contains the results obtained after applying the hybrid VNS-SA algorithm. Note that, for parameter $\eta$, its estimate value in the table corresponds to that of the
shifted process by considering $k=t_0=1980$.  The bounds obtained for $\alpha$ were $0.8724$ for scenario 1 and $0.8846$ for scenario 2.

Table \ref{tab:norway} also contains the observed values of peak time and peak, together with their respective estimates. Under scenario 1, peak estimation (\textit{Peak a)} in the table) was performed using expression (\ref{pico1}). This case allows us to validate the methodology proposed since peak was already reached in 2001. Indeed, by considering the truncated data series prior to 2001, the method predicts a peak time that is very close to the value observed. When considering scenario 2, we also included the estimation made by using (\ref{pico2}). In this latter case, value $x_s$ was the production in 1999 ($x_s=3019$ Mbbl/d) and the estimated value was noted by \textit{Peak b)} in the table. This value allows us to perform more accurate predictions regarding production peak. Note that no exact distributions are available for the ML estimators of the parameters of the model. For this reason, and with a view to provide an error value for our estimations, we considered a joint asymptotic distribution of estimators. With this as a starting point, estimation errors could be obtained for the parameters and for any parametric function, such as peak and peak time, by applying the delta method (see Appendix for more details). Table \ref{tab:norway} shows these error values below each estimation.

\begin{figure}
\caption{Observed and predicted values for Norway (scenario 1).}
\centering
  \includegraphics[width=7cm, height = 6cm]{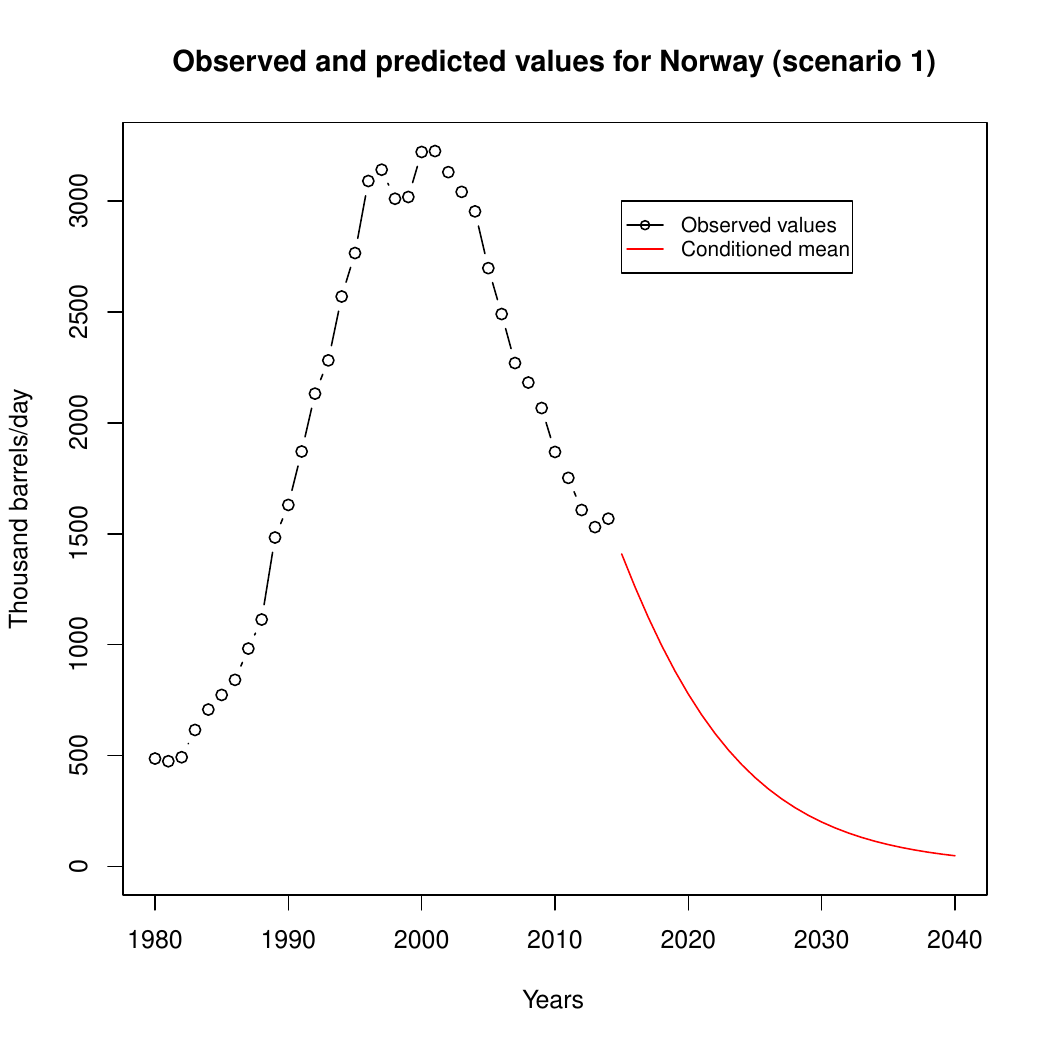}
\label{Norway1}
\end{figure}

\begin{figure}
\caption{Observed and predicted values for Norway (scenario 2).}
\centering
  \includegraphics[width=7cm, height = 6cm]{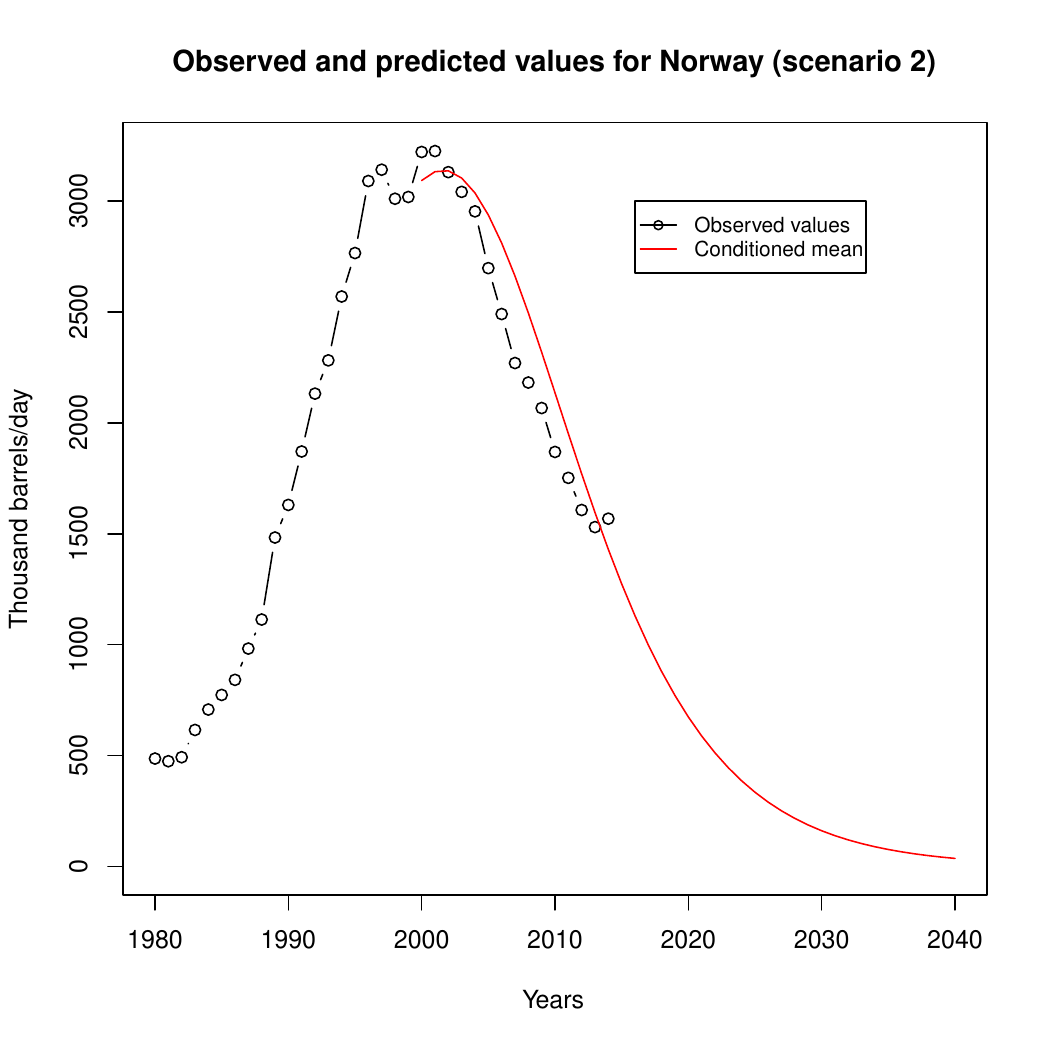}
\label{Norway2}
\end{figure}

Once the model for crude oil production has been estimated, we may infer what the production will be for the next years. In this case we have predicted production values until 2040. To this end we have considered the conditional version of the mean given by (\ref{mediacondicionada}). Figures \ref{Norway1} and \ref{Norway2} represent the observed values and the predicted ones under both scenarios. In the first case, the mean function is conditioned on the value recorded at the last observation time instant ($s=2014$, $x_s=1568$ Mbbl/d), whereas for the second case we have considered ($s=1999$, $x_s=3019$ Mbbl/d). In Figure \ref{Norway2} we have also represented the observed values until 2014 in order to visualize the overall behavior of production from the initial time of observation. Note that, after reaching its peak around 2001, Norway begins to exhibit a decline in production, and our forecasts indicate the exhaustion of its resources close to 2040 provided that a new oil cycle does not begin (something that could happen, for example, after the discovery of new resources). In addition, by using the joint asymptotic distribution of the ML estimators of the parameters and applying the delta method, asymptotic confidence intervals can be obtained for our predictions (note that the conditioned mean function is also a parametric function). Table \ref{Pred1} displays said intervals for scenario 1 (the one based on the real situation today).

\begin{figure*}
\caption{Observed and predicted values for Kazakhstan.}
\centering
  \includegraphics[width=7cm, height = 6cm]{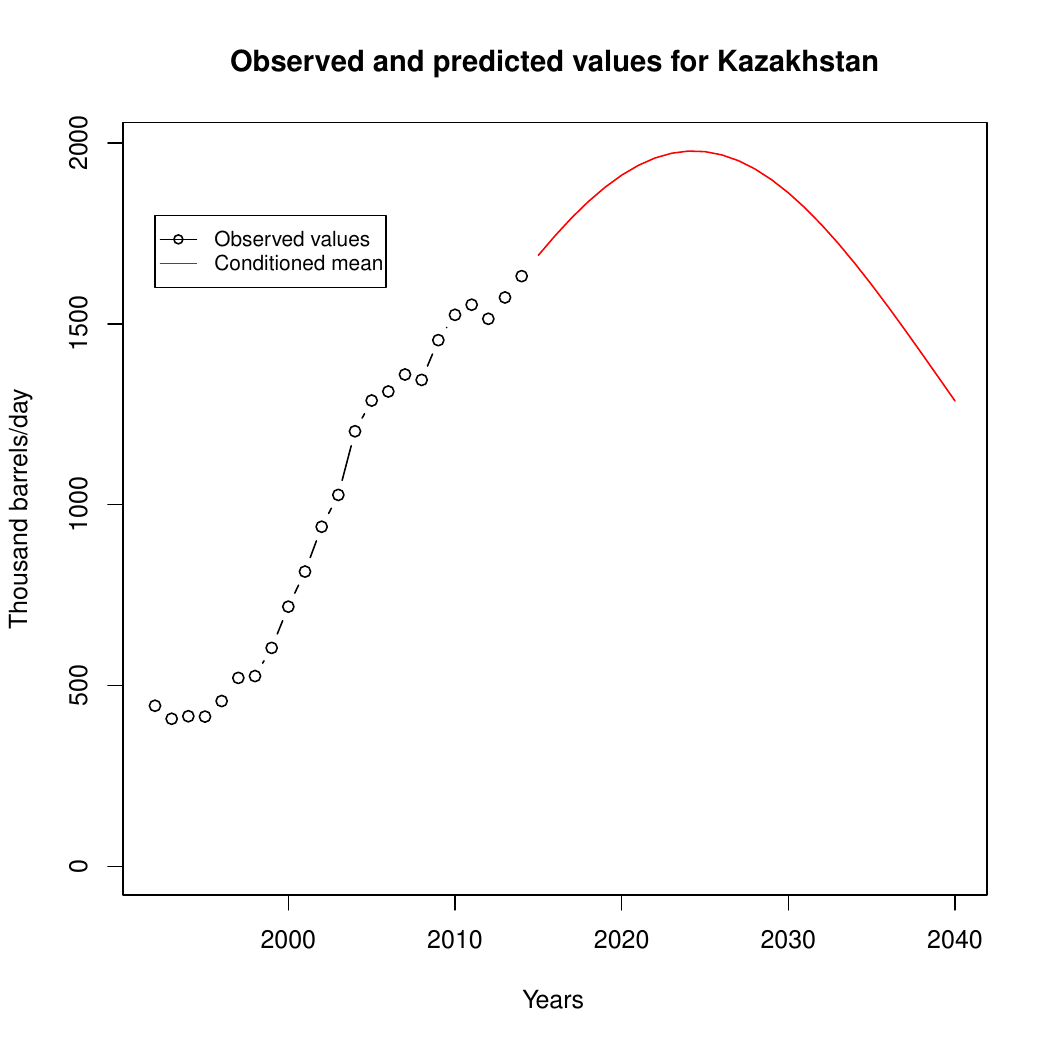}
\label{Kazakhstan}
\end{figure*}

As for Kazakhstan, it is clear that only scenario 2 can be considered, since production peak has not yet occurred. Table \ref{tab:kazakhstan} contains the results\footnote{In this case the bound of parameter $\alpha$ is 0.9603.}.
In this case, production peak was estimated by using (\ref{pico2}), considering $s=2014$ and the production in that year (1632 Mbbl/d). In a similar way to Norway, Figure \ref{Kazakhstan} shows the observed values and the forecasts made until 2040, whereas Table \ref{Pred2} contains the confidence intervals for our predictions. The results show that oil production will continue to grow until a peak time which will occur around 2025, with a predicted value of 2058.396 Mbbl/d. After that time instant, a clear decrease in production is observed, although predicted values do not yet allow to deduce the time when resources will be exhausted.

\section{Conclusions}
\label{conclusion}
A diffusion process associated with the Hubbert curve is proposed in order to study crude oil production and to predict production peak and peak time. The inferential study is carried out on the basis of discrete sampling via the maximum-likelihood method. Since a complex system of equations appears which can not be solved via classical numerical procedures, we suggest using metaheuristic optimization algorithms such as simulated annealing and variable neighborhood search to directly optimize the likelihood function. One of the fundamental problems for the application of these methods is the space of solutions, since in this case it is continuous and unbounded, which could lead to unnecessary calculation and long algorithm-running times. To this end, a strategy is suggested for bounding the space of solutions which uses the information on the characteristics of the model provided by the sample data. Simulations were performed in order to test the validity of the bounding method for the space of
solutions, showing that it is indeed very useful. The suggested bounding procedure was used in the application of the SA and VNS algorithms to estimate the parameters of the process. Both yielded good results, and the use of a hybrid VNS-SA algorithm led to substantial improvement (in terms of absolute relative errors between the estimated an the real likelihood function) when compared with the results of SA alone.

Finally, two applications to real oil production are made. Concretely, we considered oil production data from Norway and Kazakhstan. The former case allowed us to validate the procedure and methodology proposed, since peak already took place in 2001. As a matter of fact, by considering the truncated data series prior to 2001, the method predicts a peak time that is very close to the value observed in real life. Since Kazakhstani oil production has not yet reached its peak, we forecast that its growth trend will continue until a peak time which will occur with high probability around 2025. For the two countries we have considered forecasts of oil production until 2040, based on the values observed in 2014. For Norway, forecasts indicate the exhaustion of its resources at a time that is close to 2040. For Kazakhstan, a clear decrease in production is observed, although predicted values do not yet allow to deduce the time when resources will be exhausted.

\section*{Acknowledgements}
\label{acknowledgements}
This work was supported in part by the Ministerio de Econom\'ia y Competitividad, Spain, under Grant MTM2014-58061-P, and by CNPQ, Conselho Nacional de Desenvolvimento Científico e Tecnológico, Brazil.

\section*{Appendix}

This appendix summarizes the calculation of the estimation errors for the parameters of the model, as well as for several parametric functions: peak time, peak, and predictions.

In general, on the basis of the properties of the ML estimators and if $\theta$ is a multi-dimensional parameter, it is known that its ML estimator $\widehat{\theta}$ is asymptotically distributed as a normal distribution with mean $\theta$ and covariace matrix $I(\theta)^{-1}/N$, where $I(\theta)$ is the Fisher's information matrix associated with a sample of size $N$.

Accordingly, and by applying the delta method, if $g(\theta)$ is a function of the parameter, then
$$
\sqrt{N}(g(\widehat{\theta})-g(\theta)) \build{\rightarrow}{D}{} N\left[0; \nabla g(\theta)^TI(\theta)^{-1}\nabla g(\theta)\right]
$$
where $\nabla g(\theta)$ represents the derivative vector of $g(\theta)$ with respect to $\theta$.

In the problem at hand, we consider the likelihood (\ref{vero}) resulting from ignoring the initial data (therefore sample size is $N-d$) and parametric vector $\theta=(\eta,\alpha,\sigma)^T$. Thus, Fisher's information matrix is given by
$$
I(\theta)=\frac{1}{\sigma^2}
\left(
\begin{array}{ccc}
4M_1^{\eta,\alpha} & 4M_3^{\eta,\alpha}+2\frac{X_1^{\eta,\alpha}}{\alpha} &  -X_1^{\eta,\alpha}  \\ \\
4M_3^{\eta,\alpha}+2\frac{X_1^{\eta,\alpha}}{\alpha} &
4M_2^{\eta,\alpha}+\frac{Z_2}{\alpha^2}+4\frac{X_2^{\eta,\alpha}}{\alpha} &
-X_2^{\eta,\alpha}-\frac{Z_2}{2\alpha} \\ \\
-X_1^{\eta,\alpha} & -X_2^{\eta,\alpha}-\frac{Z_2}{2\alpha} & \frac{N-d}{2\sigma^2}+\frac{Z_2}{4}
\end{array}
\right)
$$
where

\begin{eqnarray*}
&M_1^{\eta,\alpha} =\displaystyle \sum_{i = 1}^d \sum^{n_i}_{j=2} \frac{\left(W_{ij}^{\eta,\alpha}/S_{ij}^{\eta,\alpha}\right)^2}{t_{ij} - t_{i,j-1}},\  M_2^{\eta,\alpha} =\displaystyle \sum_{i = 1}^d \sum^{n_i}_{j=2} \frac{\left(V_{ij}^{\eta,\alpha}/S_{ij}^{\eta,\alpha}\right)^2}{t_{ij} - t_{i,j-1}}, \
M_3^{\eta,\alpha} =\displaystyle \sum_{i = 1}^d \sum^{n_i}_{j=2} \frac{V_{ij}^{\eta,\alpha}W_{ij}^{\eta,\alpha}}{t_{ij} - t_{i,j-1}}\\
&Z_2 = \displaystyle\sum_{i = 1}^d\left(t_{in_i} - t_{i1}\right),\ X_1^{\eta,\alpha}=\displaystyle\sum_{i=1}^{d}\displaystyle\frac{W_{i}^{\alpha}}{S_{i}^{\eta,\alpha}},\
X_2^{\eta,\alpha}=\displaystyle\sum_{i=1}^{d}\displaystyle\frac{V_{i}^{\eta,\alpha}}{S_{i}^{\eta,\alpha}},\\
\end{eqnarray*}
with
\begin{eqnarray*}
&S_{ij}^{\eta,\alpha}&=(\eta + \alpha^{t_{ij-1}})(\eta + \alpha^{t_{ij}}),\ S_{i}^{\eta,\alpha} = (\eta + \alpha^{t_{i1}})(\eta + \alpha^{t_{in_i}}),\\
&W^{\alpha}_{ij}&=\alpha^{t_{ij}} - \alpha^{t_{i,j-1}},\ W^{\alpha}_{i} = \alpha^{t_{in_i}} - \alpha^{t_{i1}},\\
&V_{ij}^{\eta,\alpha}&=t_{i,j-1}\alpha^{t_{i,j-1}-1}\left(\eta+\alpha^{t_{ij}}\right)-t_{ij}\alpha^{t_{ij}-1}\left(\eta+\alpha^{t_{i,j-1}}\right),\\
&V_{i}^{\eta,\alpha}&=t_{i1}\alpha^{t_{i1}-1}\left(\eta+\alpha^{t_{in_i}}\right)-t_{in_i}\alpha^{t_{in_i}-1}\left(\eta+\alpha^{t_{i1}}\right).
\end{eqnarray*}

The elements in the diagonal of matrix $I(\theta)^{-1}/(N-d)$ provide variances for the estimations of parameters, whereas the delta method provides those of parametric functions (\ref{TiempoPico}), (\ref{pico1}), and (\ref{pico2}), which determine peak time and peak (unconditioned and conditioned). For the purpose of our predictions, function (\ref{mediacondicionada}) was used.

\newpage

\begin{table}[h]
\caption{Bounds for $\alpha$ for several values of this parameter and $\eta$.}
{\small
$$
\begin{array}{|c|c|c|c|c|c|c|c|c|}
\cline{2-9}
\multicolumn{1}{c|}{} & \multicolumn{8}{|c|}{t_0=0, \ \ t_f=50}\\
\cline{2-9}
\multicolumn{1}{c|}{} & \multicolumn{2}{|c|}{\eta=0.01} & \multicolumn{2}{|c|}{\eta=0.025} & \multicolumn{2}{|c|}{\eta=0.05} & \multicolumn{2}{|c|}{\eta=0.075}\\
\hline
\multicolumn{1}{|c|}{\alpha} & \multicolumn{1}{|c|}{\alpha_1} & \multicolumn{1}{|c|}{\alpha_2} & \multicolumn{1}{|c|}{\alpha_1} & \multicolumn{1}{|c|}{\alpha_2} & \multicolumn{1}{|c|}{\alpha_1} & \multicolumn{1}{|c|}{\alpha_2}
& \multicolumn{1}{|c|}{\alpha_1} & \multicolumn{1}{|c|}{\alpha_2} \\
\hline
0.05 & 0.8891 & 0.8088 & 0.7519 & 0.8388 & 0.5807 & 0.8619 & 0.4594 & 0.8756 \\
0.10 & 0.9136 & 0.8088 & 0.8031 & 0.8388 & 0.6585 & 0.8619 & 0.5500 & 0.8756 \\
0.15 & 0.9283 & 0.8088 & 0.8347 & 0.8388 & 0.7088 & 0.8619 & 0.6111 & 0.8756 \\
0.20 & 0.9388 & 0.8088 & 0.8579 & 0.8388 & 0.7467 & 0.8619 & 0.6584 & 0.8756 \\
0.25 & 0.9470 & 0.8088 & 0.8763 & 0.8388 & 0.7776 & 0.8619 & 0.6977 & 0.8756 \\
0.30 & 0.9538 & 0.8088 & 0.8917 & 0.8388 & 0.8037 & 0.8619 & 0.7315 & 0.8756 \\
0.35 & 0.9596 & 0.8088 & 0.9049 & 0.8388 & 0.8265 & 0.8619 & 0.7614 & 0.8756 \\
0.40 & 0.9647 & 0.8088 & 0.9164 & 0.8388 & 0.8468 & 0.8619 & 0.7883 & 0.8756 \\
0.45 & 0.9691 & 0.8088 & 0.9268 & 0.8388 & 0.8651 & 0.8619 & 0.8127 & 0.8756 \\
0.50 & 0.9731 & 0.8088 & 0.9361 & 0.8388 & 0.8818 & 0.8619 & 0.8353 & 0.8756 \\
0.55 & 0.9768 & 0.8088 & 0.9446 & 0.8388 & 0.8972 & 0.8619 & 0.8562 & 0.8756 \\
0.60 & 0.9801 & 0.8088 & 0.9525 & 0.8388 & 0.9114 & 0.8619 & 0.8758 & 0.8756 \\
0.65 & 0.9832 & 0.8088 & 0.9598 & 0.8388 & 0.9248 & 0.8619 & 0.8941 & 0.8756 \\
0.70 & 0.9861 & 0.8088 & 0.9666 & 0.8388 & 0.9373 & 0.8619 & 0.9115 & 0.8756 \\
0.75 & 0.9887 & 0.8090 & 0.9729 & 0.8388 & 0.9491 & 0.8619 & 0.9280 & 0.8756 \\
0.80 & 0.9912 & 0.8132 & 0.9789 & 0.8395 & 0.9603 & 0.8621 & 0.9437 & 0.8757 \\
0.85 & 0.9936 & 0.8546 & 0.9846 & 0.8522 & 0.9709 & 0.8660 & 0.9586 & 0.8776 \\
0.90 & 0.9958 & 0.9398 & 0.9900 & 0.9148 & 0.9810 & 0.9019 & 0.9730 & 0.9000 \\
0.95 & 0.9979 & 0.9915 & 0.9951 & 0.9821 & 0.9907 & 0.9715 & 0.9867 & 0.9644 \\
\hline
\multicolumn{1}{c|}{} & \multicolumn{2}{|c|}{\eta=0.1} & \multicolumn{2}{|c|}{\eta=0.15} & \multicolumn{2}{|c|}{\eta=0.2} & \multicolumn{2}{|c|}{\eta=0.25}\\
\hline
\multicolumn{1}{|c|}{\alpha} & \multicolumn{1}{|c|}{\alpha_1} & \multicolumn{1}{|c|}{\alpha_2} & \multicolumn{1}{|c|}{\alpha_1} & \multicolumn{1}{|c|}{\alpha_2} & \multicolumn{1}{|c|}{\alpha_1} & \multicolumn{1}{|c|}{\alpha_2}
& \multicolumn{1}{|c|}{\alpha_1} & \multicolumn{1}{|c|}{\alpha_2} \\
\hline
0.05 & 0.3714 & 0.8853 & 0.2568 & 0.8989 & 0.1893 & 0.9085 & 0.1470 & 0.9158 \\
0.10 & 0.4671 & 0.8853 & 0.3518 & 0.8989 & 0.2782 & 0.9085 & 0.2290 & 0.9158 \\
0.15 & 0.5341 & 0.8853 & 0.4228 & 0.8989 & 0.3485 & 0.9085 & 0.2969 & 0.9158 \\
0.20 & 0.5874 & 0.8853 & 0.4818 & 0.8989 & 0.4089 & 0.9085 & 0.3569 & 0.9158 \\
0.25 & 0.6323 & 0.8853 & 0.5331 & 0.8989 & 0.4629 & 0.9085 & 0.4117 & 0.9158 \\
0.30 & 0.6716 & 0.8853 & 0.5791 & 0.8989 & 0.5122 & 0.9085 & 0.4627 & 0.9158 \\
0.35 & 0.7067 & 0.8853 & 0.6210 & 0.8989 & 0.5580 & 0.9085 & 0.5107 & 0.9158 \\
0.40 & 0.7386 & 0.8853 & 0.6598 & 0.8989 & 0.6010 & 0.9085 & 0.5563 & 0.9158 \\
0.45 & 0.7679 & 0.8853 & 0.6960 & 0.8989 & 0.6417 & 0.9085 & 0.5998 & 0.9158 \\
0.50 & 0.7952 & 0.8853 & 0.7301 & 0.8989 & 0.6803 & 0.9085 & 0.6417 & 0.9158 \\
0.55 & 0.8206 & 0.8853 & 0.7624 & 0.8989 & 0.7173 & 0.9085 & 0.6820 & 0.9158 \\
0.60 & 0.8446 & 0.8853 & 0.7931 & 0.8989 & 0.7529 & 0.9085 & 0.7211 & 0.9158 \\
0.65 & 0.8672 & 0.8853 & 0.8224 & 0.8989 & 0.7871 & 0.9085 & 0.7590 & 0.9158 \\
0.70 & 0.8887 & 0.8853 & 0.8505 & 0.8989 & 0.8202 & 0.9085 & 0.7959 & 0.9158 \\
0.75 & 0.9092 & 0.8853 & 0.8776 & 0.8989 & 0.8522 & 0.9085 & 0.8318 & 0.9158 \\
0.80 & 0.9288 & 0.8854 & 0.9037 & 0.8989 & 0.8834 & 0.9085 & 0.8669 & 0.9158 \\
0.85 & 0.9476 & 0.8865 & 0.9289 & 0.8995 & 0.9136 & 0.9088 & 0.9012 & 0.9161 \\
0.90 & 0.9657 & 0.9016 & 0.9533 & 0.9078 & 0.9431 & 0.9141 & 0.9347 & 0.9198 \\
0.95 & 0.9831 & 0.9595 & 0.9769 & 0.9538 & 0.9719 & 0.9511 & 0.9677 & 0.9501 \\
\hline
\end{array}
$$
}
\label{Bounds}
\end{table}

\newpage

\begin{landscape}
\begin{table}[h]
\caption{Absolute relative errors ($\times 10^{-3}$) between the real and the estimated
log-likelihood function after applying SA and VNS-SA from simulated data.}
{\small
$$
\begin{array}{|c|r|r|r|r|r|r|r|r|r|r|r|r|r|r|r|r|}
\cline{2-17}
\multicolumn{1}{c|}{} & \multicolumn{16}{|c|}{\sigma=0.05}\\
\cline{2-17}
\multicolumn{1}{c|}{} & \multicolumn{2}{|c|}{\eta=0.01} & \multicolumn{2}{|c|}{\eta=0.025}
& \multicolumn{2}{|c|}{\eta=0.05} & \multicolumn{2}{|c|}{\eta=0.075} & \multicolumn{2}{|c|}{\eta=0.1} &
\multicolumn{2}{|c|}{\eta=0.15} & \multicolumn{2}{|c|}{\eta=0.20} & \multicolumn{2}{|c|}{\eta=0.25}\\
\hline
\multicolumn{1}{|c|}{\alpha} & \multicolumn{1}{|c|}{SA} & \multicolumn{1}{|c|}{VNS} &
 \multicolumn{1}{|c|}{SA} & \multicolumn{1}{|c|}{VNS} & \multicolumn{1}{|c|}{SA} & \multicolumn{1}{|c|}{VNS} &
 \multicolumn{1}{|c|}{SA} & \multicolumn{1}{|c|}{VNS} &
\multicolumn{1}{|c|}{SA} & \multicolumn{1}{|c|}{VNS} & \multicolumn{1}{|c|}{SA} & \multicolumn{1}{|c|}{VNS} & \multicolumn{1}{|c|}{SA} & \multicolumn{1}{|c|}{VNS} & \multicolumn{1}{|c|}{SA} & \multicolumn{1}{|c|}{VNS}\\
\hline
0.05 & 2.150 & 0.470 & 1.440 & 0.110 & 0.750 & 0.001 & 0.310 & 0.001 & 0.110 & 0.010 & 0.010 & 0.001 & 0.040 & 0.001 & 0.040 & 0.001\\
0.15 & 2.750 & 0.050 & 1.620 & 0.070 & 1.090 & 0.058 & 0.050 & 0.010 & 0.328 & 0.012 & 0.292 & 0.018 & 0.245 & 0.011 & 0.123 & 0.049\\
0.25 & 2.450 & 0.040 & 0.840 & 0.010 & 0.915 & 0.013 & 0.440 & 0.020 & 0.574 & 0.017 & 0.005 & 0.005 & 0.211 & 0.001 & 0.066 & 0.065\\
0.35 & 1.500 & 0.010 & 1.020 & 0.040 & 0.822 & 0.027 & 0.310 & 0.001 & 0.277 & 0.012 & 0.139 & 0.009 & 0.170 & 0.005 & 0.127 & 0.043\\
0.45 & 6.640 & 0.270 & 1.990 & 0.001 & 0.313 & 0.007 & 0.280 & 0.001 & 0.321 & 0.004 & 0.346 & 0.005 & 0.007 & 0.064 & 0.077 & 0.067\\
0.55 & 9.290 & 0.100 & 1.970 & 0.030 & 0.228 & 0.037 & 0.560 & 0.001 & 0.420 & 0.057 & 0.067 & 0.002 & 0.259 & 0.019 & 0.099 & 0.100\\
0.65 & 7.489 & 2.060 & 7.620 & 0.530 & 1.348 & 0.039 & 0.120 & 0.040 & 1.028 & 0.117 & 0.502 & 0.095 & 0.153 & 0.085 & 0.074 & 0.068\\
0.75 & 2.470 & 0.170 & 5.240 & 0.050 & 0.404 & 0.064 & 2.270 & 0.040 & 3.651 & 0.034 & 2.144 & 0.738 & 0.027 & 0.029 & 0.518 & 0.426\\
0.85 & 6.330 & 0.700 & 3.530 & 0.350 & 1.022 & 0.078 & 0.090 & 0.010 & 0.536 & 0.053 & 0.938 & 0.135 & 0.392 & 0.295 & 0.904 & 0.342\\
0.95 & 0.590 & 0.150 & 0.300 & 0.010 & 0.458 & 0.146 & 0.030 & 0.020 & 0.343 & 0.105 & 0.259 & 0.236 & 0.018 & 0.017 & 0.028 & 0.013\\
\hline
\multicolumn{1}{c|}{} & \multicolumn{16}{|c|}{\sigma=0.07}\\
\cline{2-17}
\multicolumn{1}{c|}{} & \multicolumn{2}{|c|}{\eta=0.01} & \multicolumn{2}{|c|}{\eta=0.025}
& \multicolumn{2}{|c|}{\eta=0.05} & \multicolumn{2}{|c|}{\eta=0.075} & \multicolumn{2}{|c|}{\eta=0.1} &
\multicolumn{2}{|c|}{\eta=0.15} & \multicolumn{2}{|c|}{\eta=0.20} & \multicolumn{2}{|c|}{\eta=0.25}\\
\hline
\multicolumn{1}{|c|}{\alpha} & \multicolumn{1}{|c|}{SA} & \multicolumn{1}{|c|}{VNS} &
 \multicolumn{1}{|c|}{SA} & \multicolumn{1}{|c|}{VNS} & \multicolumn{1}{|c|}{SA} & \multicolumn{1}{|c|}{VNS} &
 \multicolumn{1}{|c|}{SA} & \multicolumn{1}{|c|}{VNS} &
\multicolumn{1}{|c|}{SA} & \multicolumn{1}{|c|}{VNS} & \multicolumn{1}{|c|}{SA} & \multicolumn{1}{|c|}{VNS} & \multicolumn{1}{|c|}{SA} & \multicolumn{1}{|c|}{VNS} & \multicolumn{1}{|c|}{SA} & \multicolumn{1}{|c|}{VNS}\\
\hline
0.05 & 0.180 & 0.010 & 0.250 & 0.010 & 0.310 & 0.001 & 0.120 & 0.001 & 0.090 & 0.001 & 0.009 & 0.001 & 0.009 & 0.001 & 0.020 & 0.001\\
0.15 & 1.000 & 0.010 & 0.052 & 0.030 & 0.267 & 0.009 & 0.110 & 0.001 & 0.009 & 0.006 & 0.069 & 0.013 & 0.012 & 0.004 & 0.096 & 0.005\\
0.25 & 0.440 & 0.010 & 0.240 & 0.001 & 0.033 & 0.018 & 0.070 & 0.001 & 0.011 & 0.016 & 0.043 & 0.005 & 0.034 & 0.006 & 0.082 & 0.010\\
0.35 & 0.260 & 0.010 & 0.640 & 0.010 & 0.139 & 0.026 & 0.260 & 0.001 & 0.132 & 0.010 & 0.003 & 0.017 & 0.022 & 0.022 & 0.016 & 0.009\\
0.45 & 2.300 & 0.010 & 0.310 & 0.001 & 0.322 & 0.005 & 0.070 & 0.010 & 0.336 & 0.031 & 0.338 & 0.028 & 0.031 & 0.007 & 0.041 & 0.051\\
0.55 & 6.600 & 0.030 & 0.970 & 0.040 & 0.598 & 0.022 & 0.040 & 0.030 & 0.046 & 0.040 & 0.039 & 0.022 & 0.013 & 0.044 & 0.007 & 0.002\\
0.65 & 3.350 & 0.010 & 0.590 & 0.160 & 0.517 & 0.125 & 0.090 & 0.070 & 0.091 & 0.084 & 0.162 & 0.133 & 0.063 & 0.022 & 0.376 & 0.108\\
0.75 & 0.500 & 0.010 & 0.750 & 0.080 & 0.181 & 0.136 & 0.330 & 0.015 & 1.325 & 1.256 & 1.609 & 0.069 & 0.866 & 0.317 & 1.584 & 0.658\\
0.85 & 2.000 & 0.440 & 1.390 & 0.600 & 0.211 & 0.060 & 0.050 & 0.020 & 0.019 & 0.016 & 0.429 & 0.151 & 0.186 & 0.184 & 0.715 & 0.570\\
0.95 & 0.330 & 0.080 & 0.090 & 0.030 & 0.154 & 0.064 & 0.051 & 0.021 & 0.074 & 0.039 & 0.512 & 0.183 & 0.483 & 0.356 & 0.654 & 0.595\\
\hline
\end{array}
$$
}
\label{Errors}
\end{table}
\end{landscape}

\newpage

%\begin{landscape}
\begin{table}[h]
\caption{Estimated values of the parameters after applying the VNS-SA algorithm from simulated data.}
{\footnotesize
$$
\begin{array}{|c|c|c|c|c|c|c|c|c|c|c|c|c|}
\cline{2-13}
\multicolumn{1}{c|}{} & \multicolumn{12}{|c|}{\sigma=0.05}\\
\cline{2-13}
\multicolumn{1}{c|}{} &  \multicolumn{3}{|c|}{\eta=0.01} & \multicolumn{3}{|c|}{\eta=0.025} & \multicolumn{3}{|c|}{\eta=0.05} & \multicolumn{3}{|c|}{\eta=0.075}\\
\hline
\multicolumn{1}{|c|}{\alpha} &
\multicolumn{1}{|c|}{\widehat{\eta}} & \multicolumn{1}{|c|}{\widehat{\alpha}} & \multicolumn{1}{|c|}{\widehat{\sigma}} &
\multicolumn{1}{|c|}{\widehat{\eta}} & \multicolumn{1}{|c|}{\widehat{\alpha}} & \multicolumn{1}{|c|}{\widehat{\sigma}} &
\multicolumn{1}{|c|}{\widehat{\eta}} & \multicolumn{1}{|c|}{\widehat{\alpha}} & \multicolumn{1}{|c|}{\widehat{\sigma}} &
\multicolumn{1}{|c|}{\widehat{\eta}} & \multicolumn{1}{|c|}{\widehat{\alpha}} & \multicolumn{1}{|c|}{\widehat{\sigma}}\\
\hline
0.05 & 0.0101 & 0.0501 &  0.0599 & 0.0248 & 0.0500 & 0.0547 & 0.0500 & 0.0501 & 0.0511 & 0.0746 & 0.0501 & 0.0509 \\
0.15 & 0.0100 & 0.1503 &  0.0526 & 0.0249 & 0.1502 & 0.0531 & 0.0504 & 0.1501 & 0.0510 & 0.0747 & 0.1501 & 0.0502 \\
0.25 & 0.0099 & 0.2501 &  0.0516 & 0.0250 & 0.2505 & 0.0513 & 0.0500 & 0.2499 & 0.0506 & 0.0745 & 0.2499 & 0.0505 \\
0.35 & 0.0101 & 0.3501 &  0.0509 & 0.0247 & 0.3500 & 0.0514 & 0.0505 & 0.3498 & 0.0512 & 0.0751 & 0.3501 & 0.0498 \\
0.45 & 0.0098 & 0.4503 &  0.0524 & 0.0249 & 0.4502 & 0.0507 & 0.0500 & 0.4500 & 0.0508 & 0.0745 & 0.4502 & 0.0497 \\
0.55 & 0.0099 & 0.5498 &  0.0509 & 0.0249 & 0.5499 & 0.0506 & 0.0500 & 0.5499 & 0.0506 & 0.0748 & 0.5502 & 0.0501 \\
0.65 & 0.0101 & 0.6501 &  0.0503 & 0.0249 & 0.6497 & 0.0510 & 0.0502 & 0.6501 & 0.0503 & 0.0744 & 0.6499 & 0.0496 \\
0.75 & 0.0101 & 0.7501 &  0.0508 & 0.0248 & 0.7500 & 0.0508 & 0.0502 & 0.7500 & 0.0501 & 0.0742 & 0.7498 & 0.0498 \\
0.85 & 0.0092 & 0.8467 &  0.0506 & 0.0238 & 0.8480 & 0.0507 & 0.0511 & 0.8499 & 0.0502 & 0.0739 & 0.8497 & 0.0496 \\
0.95 & 0.0142 & 0.9476 &  0.0503 & 0.0253 & 0.9496 & 0.0505 & 0.0521 & 0.9498 & 0.0503 & 0.0729 & 0.9491 & 0.0498 \\
\hline
\multicolumn{1}{c|}{} &  \multicolumn{3}{|c|}{\eta=0.1} & \multicolumn{3}{|c|}{\eta=0.15} & \multicolumn{3}{|c|}{\eta=0.20} & \multicolumn{3}{|c|}{\eta=0.25}\\
\hline
\multicolumn{1}{|c|}{\alpha} &
\multicolumn{1}{|c|}{\widehat{\eta}} & \multicolumn{1}{|c|}{\widehat{\alpha}} & \multicolumn{1}{|c|}{\widehat{\sigma}} &
\multicolumn{1}{|c|}{\widehat{\eta}} & \multicolumn{1}{|c|}{\widehat{\alpha}} & \multicolumn{1}{|c|}{\widehat{\sigma}} &
\multicolumn{1}{|c|}{\widehat{\eta}} & \multicolumn{1}{|c|}{\widehat{\alpha}} & \multicolumn{1}{|c|}{\widehat{\sigma}} &
\multicolumn{1}{|c|}{\widehat{\eta}} & \multicolumn{1}{|c|}{\widehat{\alpha}} & \multicolumn{1}{|c|}{\widehat{\sigma}}\\
\hline
0.05 & 0.1001   & 0.0409    & 0.0505    & 0.1499    & 0.0499    & 0.0505    & 0.0201    & 0.0502    & 0.0504    & 0.2482    & 0.0499    & 0.0505\\
0.15 & 0.1004	& 0.1499	& 0.0506	& 0.1506	& 0.1500	& 0.0506	& 0.2005	& 0.1500	& 0.0502	& 0.2476	& 0.1498	& 0.0510\\
0.25 & 0.1004	& 0.2499	& 0.0508	& 0.1505	& 0.2499	& 0.0505	& 0.2005	& 0.2498	& 0.0505	& 0.2470	& 0.2498	& 0.0503\\
0.35 & 0.1003	& 0.3498	& 0.0507	& 0.1509	& 0.3498	& 0.0504	& 0.2005	& 0.3498	& 0.0505	& 0.2491	& 0.3498	& 0.0506\\
0.45 & 0.1004	& 0.4499	& 0.0508	& 0.1505	& 0.4499	& 0.0502	& 0.1992	& 0.4497	& 0.0503	& 0.2496	& 0.4495	& 0.0502\\
0.55 & 0.0995	& 0.5498	& 0.0503	& 0.1511	& 0.5499	& 0.0504	& 0.2009	& 0.5500	& 0.0503	& 0.2476	& 0.5497	& 0.0507\\
0.65 & 0.1004	& 0.6498	& 0.0503	& 0.1505	& 0.6497	& 0.0502	& 0.2003	& 0.6497	& 0.0503	& 0.2474	& 0.6499	& 0.0504\\
0.75 & 0.1011	& 0.7501	& 0.0505	& 0.1497	& 0.7496	& 0.0504	& 0.1988	& 0.7497	& 0.0505	& 0.2470	& 0.7496	& 0.0504\\
0.85 & 0.1012	& 0.8500	& 0.0505	& 0.1511	& 0.8498	& 0.0503	& 0.1991	& 0.8495	& 0.0504	& 0.2440	& 0.8493	& 0.0502\\
0.95 & 0.1027	& 0.9500	& 0.0502	& 0.1501	& 0.9488	& 0.0503	& 0.1939	& 0.9476	& 0.0503	& 0.2285	& 0.9469	& 0.0504\\
\hline
\end{array}
$$
}
\label{Estimates1}
\end{table}
%\end{landscape}

\newpage

%\begin{landscape}
\begin{table}[h]
\caption{Estimated values of the parameters after applying the VNS-SA algorithm from simulated data (cont.)}
{\footnotesize
$$
\begin{array}{|c|c|c|c|c|c|c|c|c|c|c|c|c|}
\cline{2-13}
\multicolumn{1}{c|}{} & \multicolumn{12}{|c|}{\sigma=0.07}\\
\cline{2-13}
\multicolumn{1}{c|}{} &  \multicolumn{3}{|c|}{\eta=0.01} & \multicolumn{3}{|c|}{\eta=0.025} & \multicolumn{3}{|c|}{\eta=0.05} & \multicolumn{3}{|c|}{\eta=0.075}\\
\hline
\multicolumn{1}{|c|}{\alpha} &
\multicolumn{1}{|c|}{\widehat{\eta}} & \multicolumn{1}{|c|}{\widehat{\alpha}} & \multicolumn{1}{|c|}{\widehat{\sigma}} &
\multicolumn{1}{|c|}{\widehat{\eta}} & \multicolumn{1}{|c|}{\widehat{\alpha}} & \multicolumn{1}{|c|}{\widehat{\sigma}} &
\multicolumn{1}{|c|}{\widehat{\eta}} & \multicolumn{1}{|c|}{\widehat{\alpha}} & \multicolumn{1}{|c|}{\widehat{\sigma}} &
\multicolumn{1}{|c|}{\widehat{\eta}} & \multicolumn{1}{|c|}{\widehat{\alpha}} & \multicolumn{1}{|c|}{\widehat{\sigma}}\\
\hline
0.05 & 0.0101 & 0.0502 & 0.0711 & 0.0253 & 0.0499 & 0.0696 & 0.0501 & 0.0498 & 0.0711 & 0.0754 & 0.0501 & 0.0701 \\
0.15 & 0.0101 & 0.1499 & 0.0716 & 0.0253 & 0.1499 & 0.0696 & 0.0499 & 0.1500 & 0.0699 & 0.0751 & 0.1501 & 0.0695 \\
0.25 & 0.0101 & 0.2502 & 0.0698 & 0.0251 & 0.2498 & 0.0691 & 0.0497 & 0.2497 & 0.0701 & 0.0751 & 0.2501 & 0.0696 \\
0.35 & 0.0099 & 0.3501 & 0.0696 & 0.0251 & 0.3501 & 0.0695 & 0.0503 & 0.3500 & 0.0693 & 0.0751 & 0.3502 & 0.0701 \\
0.45 & 0.0101 & 0.4502 & 0.0699 & 0.0252 & 0.4502 & 0.0699 & 0.0499 & 0.4501 & 0.0703 & 0.0749 & 0.4501 & 0.0698 \\
0.55 & 0.0101 & 0.5506 & 0.0703 & 0.0252 & 0.5501 & 0.0704 & 0.0498 & 0.5498 & 0.0702 & 0.0745 & 0.5499 & 0.0697 \\
0.65 & 0.0103 & 0.6507 & 0.0691 & 0.0251 & 0.6502 & 0.0696 & 0.0502 & 0.6503 & 0.0698 & 0.0738 & 0.6502 & 0.0698 \\
0.75 & 0.0103 & 0.7505 & 0.0699 & 0.0254 & 0.7506 & 0.0693 & 0.0506 & 0.7505 & 0.0702 & 0.0753 & 0.7505 & 0.0698 \\
0.85 & 0.0093 & 0.8474 & 0.0702 & 0.0233 & 0.8471 & 0.0701 & 0.0496 & 0.8499 & 0.0699 & 0.0749 & 0.8498 & 0.0701 \\
0.95 & 0.0135 & 0.9493 & 0.0698 & 0.0239 & 0.9505 & 0.0695 & 0.0480 & 0.9500 & 0.0697 & 0.0739 & 0.9494 & 0.0697 \\
\hline
\multicolumn{1}{c|}{} &  \multicolumn{3}{|c|}{\eta=0.1} & \multicolumn{3}{|c|}{\eta=0.15} & \multicolumn{3}{|c|}{\eta=0.20} & \multicolumn{3}{|c|}{\eta=0.25}\\
\hline
\multicolumn{1}{|c|}{\alpha} &
\multicolumn{1}{|c|}{\widehat{\eta}} & \multicolumn{1}{|c|}{\widehat{\alpha}} & \multicolumn{1}{|c|}{\widehat{\sigma}} &
\multicolumn{1}{|c|}{\widehat{\eta}} & \multicolumn{1}{|c|}{\widehat{\alpha}} & \multicolumn{1}{|c|}{\widehat{\sigma}} &
\multicolumn{1}{|c|}{\widehat{\eta}} & \multicolumn{1}{|c|}{\widehat{\alpha}} & \multicolumn{1}{|c|}{\widehat{\sigma}} &
\multicolumn{1}{|c|}{\widehat{\eta}} & \multicolumn{1}{|c|}{\widehat{\alpha}} & \multicolumn{1}{|c|}{\widehat{\sigma}}\\
\hline
0.05 & 0.1003   & 0.0502 & 0.0704   & 0.1507 & 0.0499   & 0.0694 & 0.2009   & 0.0501 & 0.0697   & 0.2501 & 0.0499   & 0.0701\\
0.15 & 0.0998	& 0.1500 & 0.0699	& 0.1500 & 0.1500	& 0.0692 & 0.1988	& 0.1500 & 0.0701	& 0.2468 & 0.1499	& 0.0695\\
0.25 & 0.1000	& 0.2501 & 0.0699	& 0.1504 & 0.2500	& 0.0698 & 0.2001	& 0.2498 & 0.0700	& 0.2479 & 0.2501	& 0.0699\\
0.35 & 0.1001	& 0.3501 & 0.0699	& 0.1499 & 0.3501	& 0.0697 & 0.2004	& 0.3502 & 0.0698	& 0.2466 & 0.3501	& 0.0702\\
0.45 & 0.1005	& 0.4503 & 0.0702	& 0.1509 & 0.4502	& 0.0698 & 0.2006	& 0.4500 & 0.0699	& 0.2449 & 0.4503	& 0.0701\\
0.55 & 0.0999	& 0.5502 & 0.0701	& 0.1509 & 0.5502	& 0.0703 & 0.1993	& 0.5503 & 0.0700	& 0.2479 & 0.5499	& 0.0700\\
0.65 & 0.0999	& 0.6501 & 0.0699	& 0.1510 & 0.6505	& 0.0697 & 0.2001	& 0.6501 & 0.0700	& 0.2474 & 0.6502	& 0.0698\\
0.75 & 0.1007	& 0.7505 & 0.0700	& 0.1512 & 0.7503	& 0.0699 & 0.2002	& 0.7503 & 0.0699	& 0.2458 & 0.7501	& 0.0700\\
0.85 & 0.1001	& 0.8503 & 0.0697	& 0.1505 & 0.8506	& 0.0699 & 0.1999	& 0.8504 & 0.0698	& 0.2437 & 0.8500	& 0.0699\\
0.95 & 0.0980	& 0.9499 & 0.0700	& 0.1490 & 0.9489	& 0.0701 & 0.1950	& 0.9473 & 0.0701	& 0.2488 & 0.9470	& 0.0701\\
\hline
\end{array}
$$
}
\label{Estimates2}
\end{table}
%\end{landscape}

\newpage
\begin{table}[h]
\caption{Estimated values and standard errors for Norway, considering both scenarios.}
\label{tab:norway}
{\small
$$
\begin{array}{cccccccccc}
\hline
\multicolumn{4}{c}{} & \multicolumn{2}{c}{\mbox{Observed values}} & \multicolumn{1}{c}{} & \multicolumn{3}{c}{\mbox{MLE}}\\
\cline{5-6} \cline{7-10}
\multicolumn{1}{c}{\mbox{Scenario}} & \multicolumn{1}{c}{\widehat{\eta}} & \multicolumn{1}{c}{\widehat{\alpha}} & \multicolumn{1}{c}{\widehat{\sigma}}
& \multicolumn{1}{c}{\mbox{Peak-time}} & \multicolumn{1}{c}{\mbox{Peak}} & \multicolumn{1}{c}{} & \multicolumn{1}{c}{\mbox{Peak-time}} & \multicolumn{1}{c}{\mbox{Peak a)}} & \multicolumn{1}{c}{\mbox{Peak b)}}\\
\hline
1 &  \mbox{.}0407 & \mbox{.}8638 & \mbox{.}0634 & 2001 & 3226 & & 2001\mbox{.}858 & 3226\mbox{.}144 & ----	\\
  & (.00216) & (.00177) & (.00016) & & & & (.0003) & (.40167) & \\
\hline
2 &  \mbox{.}0393 & \mbox{.}8607 & \mbox{.}0731 & 2001 & 3226 & & 2001\mbox{.}579 & 3338\mbox{.}133 & 3133\mbox{.}323  \\
  & (.00355) & (.00510) & (.00039) & & & & (.00418) & (1.01908) & (.33548)\\
\hline
\end{array}
$$
}
\end{table}

\begin{table}[h]
\caption{Estimated values and standard errors for Kazakhstan.}
\label{tab:kazakhstan}
$$
\begin{array}{ccccc}
\hline
\multicolumn{3}{c}{} & \multicolumn{2}{c}{\mbox{MLE}}\\
\cline{4-5}
\multicolumn{1}{c}{\widehat{\eta}} & \multicolumn{1}{c}{\widehat{\alpha}} & \multicolumn{1}{c}{\widehat{\sigma}}
& \multicolumn{1}{c}{\mbox{Peak-time}} & \multicolumn{1}{c}{\mbox{Peak}}\\
\hline
\mbox{.}0563 & \mbox{.}9173 & \mbox{.}0646 & 2025\mbox{.}413 & 2058\mbox{.}396  \\
(.00910) & (.00556) & (.00026) & (.03383) & (2.27192) \\
\hline
\end{array}
$$
\end{table}

\begin{table}[h]
\caption{Predicted values and confidence intervals of oil production for Norway from 2015 to 2040 when considering scenario 1.}
\label{Pred1}
\begin{center}
{
$$
\begin{array}{rrrr}
\hline
\mbox{Year} & \mbox{Mean} & \mbox{Lower Limit} & \mbox{Upper Limit} \\
\hline
2015 & 1409.457 & 1409.443 &1409.471\\
2016 & 1260.912 & 1260.887 &1260.938\\
2017 & 1123.215 & 1123.181 &1123.250\\
2018 &  996.756 &  996.715 & 996.797\\
2019 &  881.552 &  881.506 & 881.598\\
2020 &  777.338 &  777.289 & 777.387\\
2021 &  683.641 &  683.591 & 683.691\\
2022 &  599.847 &  599.796 & 599.897\\
2023 &  525.254 &  525.204 & 525.304\\
2024 &  459.119 &  459.070 & 459.168\\
2025 &  400.687 &  400.641 & 400.734\\
2026 &  349.217 &  349.173 & 349.262\\
2027 &  303.999 &  303.957 & 304.041\\
2028 &  264.363 &  264.324 & 264.402\\
2029 &  229.689 &  229.653 & 229.725\\
2030 &  199.408 &  199.374 & 199.441\\
2031 &  173.001 &  172.970 & 173.032\\
2032 &  150.004 &  149.975 & 150.032\\
2033 &  129.997 &  129.971 & 130.023\\
2034 &  112.609 &  112.586 & 112.633\\
2035 &   97.510 &   97.489 &  97.531\\
2036 &   84.407 &   84.388 &  84.427\\
2037 &   73.044 &   73.027 &  73.062\\
2038 &   63.196 &   63.180 &  63.211\\
2039 &   54.663 &   54.649 &  54.677\\
2040 &   47.274 &   47.261 &  47.286\\
\hline
\end{array}
$$
}
\end{center}
\end{table}

\begin{table}[h]
\caption{Predicted values and confidence intervals of oil production for Kazakhstan from 2015 to 2040.}
\label{Pred2}
\begin{center}
{
$$
\begin{array}{rrrr}
\hline
\mbox{Year} & \mbox{Mean} & \mbox{Lower Limit} & \mbox{Upper Limit} \\
\hline
2015 & 1694.600 & 1694.375 & 1694.826\\
2016 & 1754.204 & 1753.716 & 1754.692\\
2017 & 1810.137 & 1809.350 & 1810.925\\
2018 & 1861.734 & 1860.610 & 1862.858\\
2019 & 1908.348 & 1906.852 & 1909.844\\
2020 & 1949.373 & 1947.471 & 1951.275\\
2021 & 1984.255 & 1981.917 & 1986.593\\
2022 & 2012.508 & 2009.708 & 2015.308\\
2023 & 2033.729 & 2030.446 & 2037.012\\
2024 & 2047.610 & 2043.830 & 2051.391\\
2025 & 2053.948 & 2049.662 & 2059.233\\
2026 & 2052.647 & 2047.855 & 2057.438\\
2027 & 2043.727 & 2038.436 & 2049.017\\
2028 & 2027.320 & 2021.545 & 2033.096\\
2029 & 2003.667 & 1997.429 & 2009.906\\
2030 & 1973.110 & 1966.437 & 1979.782\\
2031 & 1936.080 & 1929.008 & 1943.152\\
2032 & 1893.088 & 1885.657 & 1900.519\\
2033 & 1844.708 & 1836.963 & 1852.454\\
2034 & 1791.563 & 1783.551 & 1799.574\\
2035 & 1734.305 & 1726.078 & 1742.533\\
2036 & 1673.607 & 1665.215 & 1681.998\\
2037 & 1610.139 & 1601.635 & 1618.642\\
2038 & 1544.561 & 1535.996 & 1553.126\\
2039 & 1477.511 & 1468.933 & 1486.088\\
2040 & 1409.589 & 1401.045 & 1418.132\\
\hline
\end{array}
$$
}
\end{center}
\end{table}

\end{document}